\documentclass[final,5p,times,twocolumn]{elsarticle}
\usepackage{lineno,hyperref}
\modulolinenumbers[5]

\usepackage{color}

\hypersetup{
	colorlinks,
	linkcolor={blue!80},
	citecolor={blue},
	urlcolor={blue!80!black}
}

\begin{document}
	
	\begin{frontmatter}
		

    \title{Spatial localization in the FitzHugh-Nagumo model}
		
		
\author{Pedro Parra-Rivas$^{a}$\corref{cor1}}
		\ead{pedro.parra-rivas@ual.es}
		\cortext[cor1]{Corresponding author}
		\affiliation{organization={Applied Physics, Department of Chemistry and Physics, University of Almeria},
			city={Almeria},
			postcode={04120}, 
			country={Spain}}
		
\author{Fahad Al Saadi$^b$}
				\affiliation{organization={Department of Systems Engineering, Military Technological College},
			city={Muscat},
			country={Oman}}
		
\author{Lendert Gelens$^c$}
		\affiliation{organization={ Laboratory of Dynamics in Biological Systems, KU Leuven, Department of Cellular and Molecular Medicine,University of Leuven}, 
			city={Leuven},
				postcode={B-3000}, 
			country={Belgium}}

		\begin{abstract}
The FitzHugh-Nagumo model, originally introduced to study neural dynamics, has since found applications across diverse fields, including cardiology and biology. However, the formation and bifurcation structure of spatially localized states in this model remain underexplored. In this work, we present a detailed bifurcation analysis of such localized structures in one spatial dimension in the FitzHugh-Nagumo model. We demonstrate that these localized states undergo a smooth transition between standard and collapsed homoclinic snaking as the system shifts from pattern-uniform to uniform-uniform bistability. Additionally, we explore the oscillatory dynamics exhibited by these states when varying the time-scale separation and diffusion coefficient. Our study leverages a combination of analytical and numerical techniques to uncover the stability and dynamic regimes of spatially localized structures, offering new insights into the mechanisms governing spatial localization in this widely used model system.
		\end{abstract}
		
		
	\end{frontmatter}
	

\section{Introduction}

Since its proposal more than sixty years ago for studying neural behavior \cite{1,2}, the FitzHugh-Nagumo (FHN) model has been extensively used to investigate a wide variety of temporal and spatiotemporal processes across disciplines ranging from cardiology to biology \cite{cebrian-lacasa_six_2024}. Despite its broad applicability, the formation of non-traveling localized states (hereafter LSs) in this model has received significantly less attention compared to its other uses.



Spatial localization in systems far from thermodynamic equilibrium arises from a dual balance: on the one hand, the interplay between nonlinearity and spatial coupling (e.g., diffusion), and on the other, the continuous exchange of energy with the surroundings \cite{akhmediev_dissipative_2005}. This framework underpins the emergence of localized states in various reaction-diffusion systems, where self-organizing processes often lead to the formation of spatially structured patterns, as initially described in Turing's seminal theory of morphogenesis~\cite{Turing}. However, in the context of the FHN model, while some studies have shown the formation of specific types of LSs and so-called homoclinic snaking \cite{cebrian-lacasa_six_2024,frohoff2023stationary}, fundamental questions regarding their origin, bifurcation structure, and stability across system parameters remain unresolved.

Answering these questions begins with an analysis of the simplest bifurcation scenario in a single extended dimension, that is, the one-dimensional (1D) case. The 1D configuration not only simplifies the mathematical framework, but also provides crucial insights into mechanisms that govern more complex spatial dynamics in higher dimensions. Therefore, this work focuses on conducting a detailed bifurcation analysis of the 1D FHN model. This includes exploring localized states alongside uniform states and spatially periodic patterns, as well as dynamic regimes like oscillatory behavior.

To achieve this, we use a combination of analytical and numerical methods widely established in pattern-forming systems. These approaches allow us to address the transitions between different localization regimes, characterize their stability, and uncover connections to broader phenomena such as Turing instabilities and front dynamics~\cite{homburg_chapter_2010}. By extending the understanding of LSs in the FHN model, we provide a foundation for exploring richer spatial phenomena in higher-dimensional settings.

This article is organized as follows. In Section~\ref{sec:1}, we introduce the FitzHugh-Nagumo model and outline the main methodological framework used in our study. Section~\ref{sec:2} focuses on the uniform extended states, including an analysis of their spectral stability. In Section~\ref{sec:3}, we derive approximate solutions for weakly nonlinear patterns and LSs using analytical techniques. Section~\ref{sec:4} presents a detailed bifurcation analysis of static LSs in the absence of dynamical instabilities. Subsequently, in Section~\ref{sec:5}, we investigate the emergence and stability of two distinct types of oscillatory LSs. In Section~\ref{sec:6}, we provide a complete stability map of the system. Finally, Section~\ref{sec:7} concludes with a discussion of our findings and a summary of the main conclusions.

\section{The model and mathematical framework}\label{sec:1}
The form of the FHN equation used in this work is:
\begin{equation}\label{model}
    \begin{array}{l}
        \partial_t u = \delta^2 \nabla^2 u + u - u^3 - v, \\ \\  
        \partial_t v = \nabla^2 v + \varepsilon (u - \alpha v - \beta),
    \end{array}
\end{equation}
where $\nabla^2$ is the Laplacian operator, modeling diffusion in general. The parameter $\delta^2$ represents the ratio of diffusion coefficients between the fields $u$ and $v$, while $\varepsilon > 0$ determines the time-scale separation between the two equations. The parameters $\alpha, \beta \in \mathbb{R}$ control the nonlinearity of the system.  

In this work, we focus on a 1D configuration. Consequently, we set $\nabla^2 \equiv \partial^2_x$ in all subsequent analyses.

\subsection{The time-independent problem and the spatial dynamics formulation}
In this work, we are particularly interested in time-independent, or steady-state, solutions, $(u,v)=(u_s,v_s)$ (i.e., $\partial_t u_s=\partial_t v_s=0$) of model (\ref{model}). These states satisfy the equation 
\begin{equation}\label{sta_eq}
\begin{array}{l}
\delta^2 \nabla^2 u+u-u^3-v=0,\\	\\
\nabla^2 v+\varepsilon(u-\alpha v-\beta)=0,
\end{array}
\end{equation}
Time-independent states of this model include spatially extended solutions, which can be uniform or not (i.e., patterns), and spatially localized states. Here, we will mainly focus on the latter, although we must first acquire a solid understanding of the former. 
	
These states can also be described using a spatial dynamical formulation of our time-independent problem \cite{haragus_local_2011}. This approach consists of writing the stationary equation (\ref{sta_eq}) as a dynamical system where the role of time is now played by $x$. This leads to the 4D dynamical system:
\begin{equation}\label{SD_compact}
\frac{d\vec{y}}{dx}=\vec{f}\left(\vec{y};\alpha,\beta,\delta,\varepsilon\right),
\end{equation}
with the vector field $\vec{f}(x)\equiv\left[f_1(\vec{y}),f_2(\vec{y}),f_3(\vec{y}),f_4(\vec{y})\right]^T$ defined by:
\begin{equation}\begin{array}{l}\label{SD_full}
f_1=y_3,\\
f_2=y_4,\\
f_3=\delta^{-2}(y_1^3+y_2-y_1),\\
f_4=\varepsilon(\alpha y_2-y_1+\beta),
\end{array}\end{equation}
and the new variables 
$\vec{y}(x)\equiv\left[y_1(x),y_2(x),y_3(x),y_4(x)\right]^T$, where $y_1(x)=u(x)$, $y_2(x)=v(x)$, $y_3(x)=du/dx$, and $y_4(x)=dv/dx$.
	
In this context, any time-independent state of the original equation has a counterpart in a 4D phase space. With this duality, the fundamental extended states of the system—that is, a uniform or homogeneous steady state and a spatially periodic pattern—correspond to a fixed point and a limit cycle of Eq.~(\ref{SD_compact}). Similarly, localized states bi-asymptotic to a uniform state are dual to homoclinic orbits leaving and approaching a fixed point \cite{parra-rivas_organization_2023}.

\subsection{The methodology}
The investigation we present here applies several techniques commonly used in the nonlinear domain. These include analytical methods, such as multiscale perturbation theory, which allows us to reduce our original model to normal forms in the weakly nonlinear regime \cite{parra-rivas_organization_2023}, and numerical algorithms for solving Eq.~(\ref{model}) in the highly nonlinear regime.

Numerically, Eq.(\ref{model}) can be solved through two main approaches. First, by considering an initial value problem and studying how initial conditions evolve in time. This allows us to analyze transient and chaotic dynamics of the system. To do so, we use pseudo-spectral methods \cite{bibid}. Second, we use path-continuation algorithms based on the Newton-Raphson method to compute and determine the bifurcation structure of either static states (e.g., patterns, LSs) or periodic temporal oscillations (e.g., oscillons). For static states, we apply the software AUTO-07p to Eq.(\ref{SD_compact}) \cite{doedel_numerical_1991} or pde2path directly to Eq.~(1) \cite{uecker_numerical_2021}. In both cases, the analytical solutions obtained through weakly nonlinear analysis are used as initial guesses for this numerical procedure. To compute the bifurcation structure and stability of oscillatory states, we also use pde2path.

After computing the bifurcation diagrams associated with the time-independent states $u_s$ ($v_s$), we perform their spectral stability analysis by solving the linear eigenvalue problem
\begin{equation} L_s\Psi=\sigma\Psi, \end{equation} where $L_s$ is the linear operator (see \ref{append1})
\begin{equation} L=L(u_s)\equiv\left[\begin{array}{cc} 1+\delta^2\partial^2_x -3u_s^2 & -1\\ \varepsilon & \partial^2_x-\varepsilon \alpha \end{array} \right], \end{equation} evaluated at the steady-state solutions $u_s$ ($v_s$). Here, $\sigma$ is the eigenvalue associated with the eigenfunction $\Psi$. The steady state is unstable if Re$[\sigma]>0$, and it evolves to a state with the form of $\Psi$.

Similarly, for time-dependent oscillatory states, we assess their stability by performing a Floquet analysis and computing the Floquet multipliers. This information is obtained simultaneously while computing these states using pde2path \cite{uecker_numerical_2021}.

    \section{The homogeneous steady state}\label{sec:2}

\begin{figure}[!t]
	\centering
	\includegraphics[scale=1]{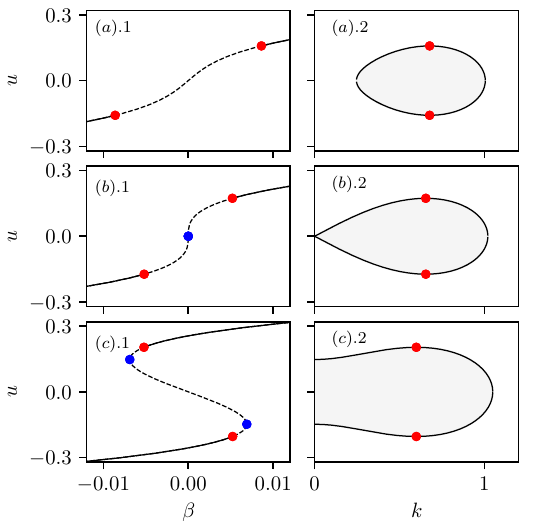}
	\caption{\textbf{Linear stability of the homogeneous steady state solutions.}
    Homogeneous steady states (left) and their corresponding linear stability analysis (right) for different values of $\alpha$. Panels (a), (b), and (c) show results for $\alpha = 0.97$, $\alpha = 1$, and $\alpha = 1.07$, respectively, with fixed parameters $\epsilon = 1$ and $\delta = 0.7$. Red dots indicate Turing instabilities, while blue dots represent saddle-node bifurcations.}
	\label{marginal}         
\end{figure}

The homogeneous steady states, or uniform solutions, of the system, $(u,v) = (U_h, V_h)$, imply $\partial_x U_h = \partial_x V_h$, which leads to the equations
\begin{equation}\label{hom_sol} \alpha U_h^3 + (1-\alpha)U_h - \beta = 0, \hspace{1cm} V_h = U_h(1 - U_h^2), \end{equation}
which are nonlinear in $U_h$. The variation of $U_h$ as a function of $\alpha$ and $\beta$ is shown in Fig.\ref{marginal}(a).1-(c).1 for specific regimes. As a function of $\beta$, $U_h$ is single-valued for $\alpha < 1$ [see Fig.\ref{marginal}(a).1]. However, for $\alpha > 1$, this equation supports three solutions, which we label $U_h^b$, $U_h^m$, and $U_h^t$ [Fig.~\ref{marginal}(c).1].

These states are separated by folds or turning points, which occur at
\begin{equation}\label{folds_eq} U_h = U_f^{l,r} \equiv \pm\sqrt{\frac{\alpha-1}{3\alpha}}. \end{equation}
At $\alpha = 1$, these two folds collide and disappear in a cusp bifurcation $C_h$ [Fig.~\ref{marginal}(b).1].

\begin{figure*}[!t]
\centering
\includegraphics[scale=1]{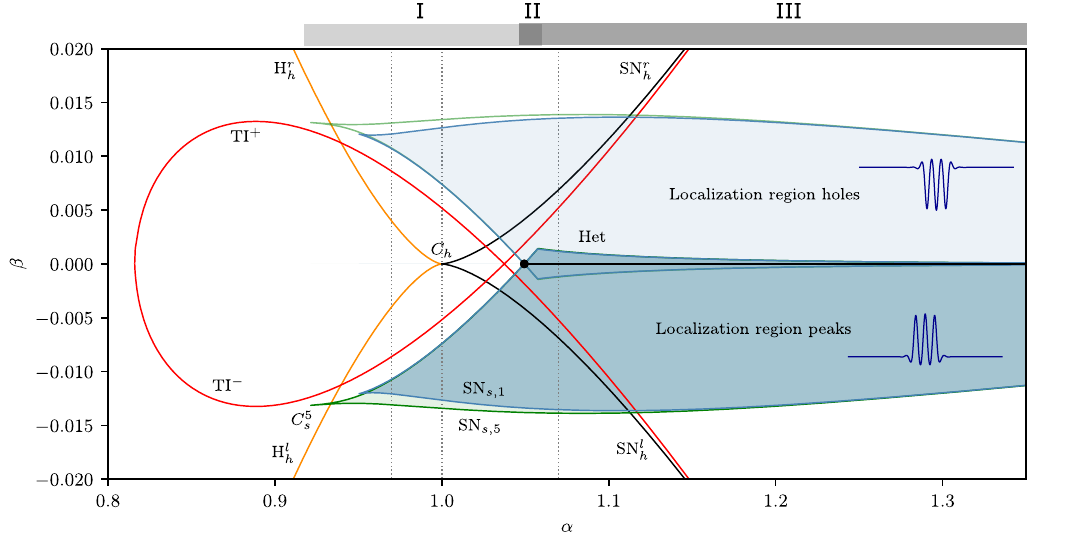}
\caption{\textbf{Phase diagram in the $(\alpha,\beta)$ parameter space.} Parameters $\epsilon = 1$ and $\delta = 0.7$ are kept fixed and the diagram shows the main bifurcation curves of the system. These include the Turing instability (TI) (red), the saddle-node bifurcations of the uniform state SN$h^{l,r}$ (black), the Hopf bifurcation of the uniform states H$h$, and the saddle-node bifurcations SN${1}^{l,r;t,b}$ that delineate the localization regions. The diagram also highlights the cusp bifurcation $C_h$ of the uniform state, the points $C{b,t}$, the degenerate Turing instabilities codim-2 bifurcations TI$^d_{1,2}$, and the codimension-2 heteroclinic point Het$_p$.
Within each localization region, three distinct sectors are identified: I, featuring standard homoclinic snaking; II, characterized by collapsed snaking; and III, representing the transition zone. Additionally, a region of purely uniform oscillatory behavior and the Turing-Hopf dynamical sector are marked.}\label{fig_phase_dia_1}         
	\end{figure*}

The equation for the folds can also be written as
\begin{equation} \beta_{f}^{l,r} = \pm\frac{2}{3}(1-\alpha)\sqrt{\frac{\alpha-1}{3\alpha}}. \end{equation}
Using this expression, we can represent how the folds vary in the $(\alpha, \beta)$ parameter space. This variation is depicted in the phase diagram shown in Fig.~\ref{fig_phase_dia_1} (see black line).

\subsection{Linear stability analysis of the flat solutions}

In most cases, we do not have an analytical expression for the steady-state solutions, and this problem must be solved numerically.

The stability of the HSS can, however, be determined analytically. This analysis provides information about how the HSS behaves against perturbations proportional to $(u,v)^T \propto \psi_k e^{\sigma t}$, where $\psi_k = e^{ikx}$.

This analysis leads to the perturbation growth equation  
\begin{equation}\label{characteristic}
	\sigma^2 - T_1(k)\sigma + \Delta_1(k) = 0,
\end{equation}
with 
\begin{equation}
	T_1(k) \equiv -(\delta^2 + 1)k^2 - 3U_h^2 - \varepsilon \alpha + 1,
\end{equation}
and 
\begin{equation}
	\Delta_1(k) \equiv (3U_h^2 + \delta^2k^2 - 1)(\varepsilon\alpha + k^2) + \varepsilon.
\end{equation}

The solution of Eq.~(\ref{characteristic}) gives the dispersion relation 
\begin{equation}\label{disperelation}
	\sigma(k) = \frac{1}{2}\left(-T_1(k) \pm \sqrt{T_1(k)^2 - 4\Delta_1(k)}\right),
\end{equation}
and different instabilities may occur:
\begin{itemize}
	\item If $Im[\sigma(k)] = 0$ and $Re[\sigma(k)] = 0$ at $k = 0$, the HSS solution undergoes a {\it saddle-node} bifurcation.
	\item If $Im[\sigma(k)] \neq 0$ and $Re[\sigma(k)] = 0$ at $k = 0$, the flat solution undergoes a {\it Hopf} bifurcation.
	\item If $Im[\sigma(k)] = 0$ and $Re[\sigma(k)] = 0$ at $k = k_T$, it undergoes a {\it Turing} bifurcation.
	\item If $Im[\sigma(k)] \neq 0$ and $Re[\sigma(k)] = 0$ at $k = k_w$, it undergoes a {\it wave} bifurcation, also called {\it oscillatory Turing}.
\end{itemize}

    \subsubsection{Saddle-node instability}
If $Im[\sigma(k)]=0$ and $Re[\sigma(k)]=0$ at $k=0$, the HSS solution undergoes a {\it saddle-node} bifurcation. These conditions imply that $\Delta_1(0)=0$, which leads to the fold positions given by Eq.~(\ref{folds_eq}). Therefore, at each of the folds, a saddle-node bifurcation takes place. 

\subsubsection{Hopf instability}
The system undergoes a Hopf instability if $Im[\sigma(k)]\neq 0$ and $Re[\sigma(k)]=0$ at $k=0$. These conditions imply $T_1(0)=0$ and $\Delta_1(0)>0$. The first condition leads to the Hopf threshold (i.e., determines the parameter values at which the Hopf bifurcation takes place):
\begin{equation}
    U_h^H=\pm\sqrt{\frac{1-\varepsilon \alpha}{3}},
\end{equation}
while the second one determines the frequency of the oscillations emerging from this point:
\begin{equation}
    \sigma=\pm i\omega_H=\pm i\sqrt{\Delta_1(0)}=\pm i\sqrt{\varepsilon-\varepsilon^2\alpha^2}.
\end{equation}
The Hopf bifurcation exists whenever $\varepsilon \alpha^2<1$ and disappears at $\alpha=1/\varepsilon$. The modification of this line is depicted in Fig.~\ref{fig_phase_dia_1} for $\varepsilon=1$. For this value, the homogeneous oscillations of the system cease precisely at the cusp bifurcation $C_h$, where bistability appears.
\subsubsection{Turing instability}
The Turing bifurcation occurs when $\sigma=0$, which is equivalent to the condition $\Delta_1(k)=0$. This condition leads to  
\begin{equation}\label{MIC}
U^2_\pm=\frac{1}{3}\left(1-\delta^2k^2-\frac{\varepsilon}{\varepsilon\alpha+k^2}\right),
\end{equation}
which defines the boundary of the HSS unstable region to perturbations with wavelength $k$. This curve is known as the marginal or neutral stability curve. The modification of this curve and the associated stability regime is shown in Figs.~\ref{marginal}(a).2-(b).3 for $\delta=0.7$, $\varepsilon=1$, and the same $\alpha$-values as the $U_h$ states on the left.

This curve has two extrema (a maximum and a minimum) occurring at $k=k_T$, which corresponds to the Turing instability \cite{cross_pattern_1993}. This condition leads to 
\begin{equation}\label{kc}
k_T=\sqrt{-\varepsilon \alpha+\sqrt{\varepsilon}/\delta},
\end{equation}
provided that $\delta<1/(\alpha\sqrt{\varepsilon})$.  
Inserting this expression into Eq.~(\ref{MIC}), we obtain the exact position of the Turing instability:  
\begin{equation}
U^{\pm}_T=\pm\sqrt{\frac{1+\delta^2\varepsilon\alpha-2\delta\sqrt{\varepsilon}}{3}}. 
\end{equation}
The Turing instability is marked using a red dot in both the $u(\beta)$ and $u(k)$ curves.

In-between these points, the HSS is unstable to spatially modulated perturbations with wavenumbers inside the gray-shaded region in Fig.~\ref{marginal}(a).2-(c).2. This is indicated using dashed lines in Fig.~\ref{marginal}(a).1-(c).1. The solid parts of these curves represent the HSS that are unstable (i.e., they lie outside the unstable regions on the right).

The modification of the Turing instability in the $(\alpha,\beta)$-plane is given by the expression  
\begin{equation}
\beta^{\pm}_T=\frac{1}{3}\left(\varepsilon\delta^2\alpha^2-2(1+\delta\sqrt{3})\alpha+3\right)U_T^\pm,
\end{equation}
and is illustrated using a red solid line in Fig.~\ref{fig_phase_dia_1}.


\section{Bifurcation in the spatial dynamics context: insights for spatial localization}

The results discussed previously refer to temporal stability. However, additional insights can be obtained by studying the spatial stability of the system by analyzing the dynamical system (\ref{SD_compact}). In particular, the analogy between LSs and homoclinic orbits allows us to predict, using spatial stability, the types of LSs that arise around the different bifurcations of the HSSs \cite{woods_heteroclinic_1999,haragus_local_2011,parra-rivas_formation_2020,parra-rivas_origin_2021,parra-rivas_organization_2023}. This information is encoded in the spectrum ${\lambda_i}$ with $i \in {1,\dots,4}$ of the Jacobian $\mathcal{J}$ associated with Eq.~(\ref{SD_compact}), which satisfies the characteristic equation
\begin{equation}
{\rm det}\left(\mathcal{J}-\lambda\mathbf{I}_{4\times4}\right)=0.
\end{equation}
This leads to the characteristic polynomial
\begin{equation}\label{eigenvalues}
\delta^2\lambda^4-(3U_h^2+\delta^2\varepsilon\alpha-1)\lambda^2+(3U_h^2-1)\varepsilon\alpha+\varepsilon=0,
\end{equation}
which can also be derived from the condition $\Delta_1(-i\lambda)=0$. Equation~(\ref{eigenvalues}) is invariant under the transformations $\lambda\rightarrow-\lambda$ and $\lambda\rightarrow\lambda^*$, resulting in eigenvalue configurations symmetric with respect to both axes. Depending on the parameter regime, different types of LSs may bifurcate from either the folds or the Turing instability.

In the spatial dynamics context, the Turing instability corresponds to a Hamiltonian-Hopf (HH) bifurcation of the dynamical system (\ref{SD_compact}). This bifurcation is defined by the degenerate eigenspectrum
\begin{equation}
\lambda_{1,2}=i\sqrt{|\varepsilon\alpha-\sqrt{\varepsilon}/\delta|},\quad \lambda_{3,4}=-i\sqrt{|\varepsilon\alpha-\sqrt{\varepsilon}/\delta|},
\end{equation}
provided that $\varepsilon\alpha-\sqrt{\varepsilon}/\delta<0$, which is equivalent to $\delta<1/(\alpha\sqrt{\varepsilon})$. Several studies have demonstrated that families of wild homoclinic orbits, i.e., LSs with oscillatory tails, emerge subcritically from this bifurcation \cite{woods_heteroclinic_1999,haragus_local_2011,champneys_homoclinic_1998}.

When $\delta>1/(\alpha\sqrt{\varepsilon})$, the HH bifurcation transitions into a Belyakov-Devaney (BD) bifurcation, where LSs are destroyed in a complex process involving Shilnikov homoclinic orbits \cite{parra-rivas_bifurcation_2018,verschueren_dissecting_2021}. This transition is characterized by the eigenspectrum
\begin{equation}
\lambda_{1,2}=\sqrt{\varepsilon\alpha-\sqrt{\varepsilon}/\delta},\quad \lambda_{3,4}=-\sqrt{\varepsilon\alpha-\sqrt{\varepsilon}/\delta}.
\end{equation}

At the folds, the spatial eigenspectra are given by
\begin{equation}
\lambda_{1,2}=0,\quad \lambda_{3,4}=\sqrt{\frac{\varepsilon\alpha^2\delta^2-1}{\delta^2\alpha}},
\end{equation}
and lead to two different bifurcations depending on the sign of $\varepsilon\alpha^2\delta^2-1$. When $\varepsilon\alpha^2\delta^2-1<0$, $\lambda_{3,4}$ are purely imaginary, and the fold points correspond to Reversible Takens-Bogdanov-Hopf (RTBH) bifurcations \cite{champneys_homoclinic_1998}. In this scenario, generalized solitary waves, biasymptotic to a small-amplitude spatial pattern of arbitrary constant amplitude, arise \cite{haragus_local_2011}. However, these waves are typically temporally unstable and, thus, unobservable \cite{parra-rivas_organization_2023}. The RTBH bifurcation coexists with the HH bifurcation in the same parameter regime.

When $\varepsilon\alpha^2\delta^2-1>0$, $\lambda_{3,4}$ are real, and the folds correspond to Reversible Takens-Bogdanov (RTB) bifurcations. In this case, tame homoclinic orbits arise, representing LSs with monotonic tails. These solutions can lead to complex bifurcation structures \cite{parra-rivas_bifurcation_2018}. This bifurcation occurs under the same parameter regime as the BD transition.

All these scenarios converge at a quadruple-zero (QZ) codimension-two bifurcation, which occurs at the folds of the HSS when $\delta=1/(\alpha\sqrt{\varepsilon})$. Given the complexity of these scenarios, the following analysis focuses on the LSs emerging from the HH point.

\section{Small amplitude localized states: Weakly nonlinear analysis near the Turing bifurcation}\label{sec:3}

In this section, we compute, using multiscale perturbation theory, weakly nonlinear steady states of the FHN model in the vicinity of the TI.  

Following \cite{burke_classification_2008, parra-rivas_origin_2021,parra-rivas_organization_2023}, we fix the values of $\varepsilon$, $\Delta$, and $a$ and assume that the states near the bifurcation are captured by the ansatz
\begin{equation}\label{ansatz}
\left[\begin{array}{c}
	u\\v
\end{array}\right]= \left[\begin{array}{c}
	U_h\\V_h
\end{array}\right]+ \left[\begin{array}{c}
	\phi(x)\\\psi(x)
\end{array}\right],
\end{equation}
where $U_h$ and $V_h$ correspond to the HSS solution, and $\phi(x)$ and $\psi(x)$ capture the spatial dependence. We introduce appropriate asymptotic expansions for each variable in terms of a small parameter, specifically $\epsilon=\sqrt{U_h-U_T}$ for the TI. We use $\beta$ as a bifurcation parameter and thus write $\beta=\beta_T+\epsilon^2\mu$ (see \ref{append2}).  

In the neighborhood of this bifurcation, weakly nonlinear states are captured by the ansatz:
\begin{equation}
\left[\begin{array}{c}
	u(x)\\v(x)
\end{array}\right]-\left[\begin{array}{c}
	U_h\\V_h
\end{array}\right] \sim \epsilon A(X)e^{ik_Tx}+c.c.,
\end{equation}
where \(A(X)\) is the amplitude, or envelope, describing a modulation occurring at a larger scale \(X\equiv\epsilon x\). This amplitude satisfies the time-independent normal form equation:
\begin{equation}\label{normal_form}
	\mu A+C_2 A_{XX}+C_3 A|A|^2=0,
\end{equation}
with coefficients \(C_{2,3}\) depending on the main control parameters of the system (see \ref{append2}). By taking \(A(X)=Z(X)e^{i\varphi}\), with \(\varphi\neq\varphi(X)\), the previous equation reduces to:
\begin{equation}\label{normal_form2}
	\mu Z+C_2 Z_{XX}+C_3 Z^3=0.
\end{equation}
When \(Z\neq Z(X)\), the solutions of Eq.~(\ref{normal_form2}) are
\begin{equation}\label{sol_pattern}
Z=0,\qquad Z=\sqrt{-\mu/C_3},
\end{equation}
which represent different solution branches emerging from a pitchfork bifurcation occurring at \(\mu=0\). Depending on the sign of \(C_3\), this solution exists for \(\mu<0\) if \(C_3>0\), or for \(\mu>0\) if \(C_3<0\). In the first case, the system exhibits a subcritical pitchfork regime, while in the second, the pitchfork is supercritical. The transition between these two situations occurs at \(C_3=0\), corresponding to a degenerate Turing instability (dTI), which is a codimension-two point. For the parameters discussed in the previous section (see phase diagram in Fig.~\ref{fig_phase_dia_1}), this point occurs at \(\alpha_c\approx 0.894048\).

The solution (\ref{sol_pattern}) corresponds to spatially extended periodic patterns of the form:
$$\left[\begin{array}{c}
	u_P(x)\\v_P(x)
\end{array}\right]=\left[\begin{array}{c}
	U_T\\V_T
\end{array}\right]+\left(\beta-\beta_T\right)\left[\begin{array}{c}
	G_1^{(1)}\\G_2^{(2)}
\end{array}\right]$$
\begin{equation}
\qquad\qquad\quad+2\left[\begin{array}{c}
1\\L_{11}^{(1)}
\end{array}\right]\sqrt{\frac{\beta-\beta_T}{-C_3}}{\rm cos}(k_Tx+\varphi),
\end{equation}
where the definitions of \(C_3\), \(L_{11}^{(1)}\), \(G_2^{(2)}\), and \(G_2^{(1)}\) are given in \ref{append1} and \ref{append2}.

In the subcritical regime, Eq.~(\ref{normal_form2}) also supports pulse solutions of the form:
\begin{equation}
Z(X)=\sqrt{\frac{-2\mu}{C_3}}{\rm sech}\left(\sqrt{\frac{-\mu}{C_2}}X\right),
\end{equation}
which yield the small-amplitude LSs:
$$\left[\begin{array}{c}
	u_P(x)\\v_P(x)
\end{array}\right]=\left[\begin{array}{c}
	U_T\\V_T
\end{array}\right]+\left(\beta-\beta_T\right)\left[\begin{array}{c}
	G_2^{(1)}\\G_2^{(2)}
\end{array}\right]$$
\begin{equation}\label{loc_states}
+2\sqrt{\frac{2(\beta-\beta_T)}{-C_3}}\left[\begin{array}{c}
		1\\L_{11}^{(1)}
	\end{array}\right]{\rm sech}\left(\sqrt{\frac{\beta-\beta_T}{-C_2}}x\right){\rm cos}(k_Tx+\varphi).
\end{equation}

The spatial phase \(\varphi\) of the periodic states is arbitrary, reflecting invariance under translations. However, this symmetry is broken for LSs, where beyond-all-orders calculations predict two specific \(\varphi\)-values, \(\varphi=0,\pi\), both preserving the spatial reversibility symmetry \((x,u,v)\rightarrow(-x,u,v)\) of Eq.~(\ref{model}) \cite{kozyreff_asymptotics_2006}. These two \(\varphi\)-values yield two types of localized weakly nonlinear solutions: one with a maximum at the center of the domain (\(x=0\)), corresponding to \(\varphi=0\), and another with a minimum at \(x=0\), associated with \(\varphi=\pi\).

\begin{figure*}[!t]
	\centering
	\includegraphics[scale=1]{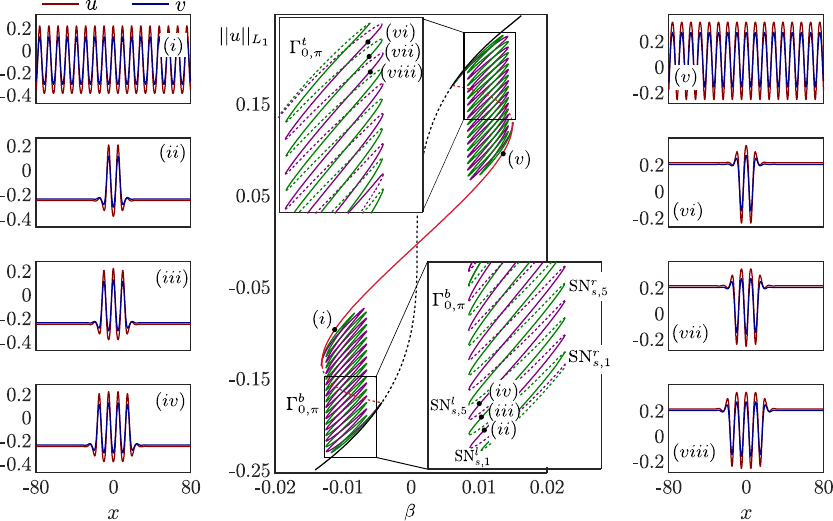}
	\caption{{\bf Standard homoclinic snaking.} This bifurcation diagram corresponds to the parameters $(\epsilon, \delta, \alpha) = (1, 0.7, 1)$. It illustrates the HSSs at the cusp (black), periodic Turing patterns emerging from TI$_{1,2}$ (red), and two groups of LSs undergoing standard snaking—one at the top and one at the bottom of the figure. Each group consists of the families $\Gamma_{0,\pi}^{b,t}$, where the superscript $b$ ($t$) denotes bottom (top) LSs. For each snaking region, a close-up view is provided, with spectral stability indicated by solid lines (stable states) and dashed lines (unstable states).  
Sample profiles of LSs and patterns along these families are shown in panels A–D for the bump states and panels E–H for the hole states.}
	\label{fig_standard}         
\end{figure*}

\section{Bifurcation structure of static localized states}\label{sec:4}

The small-amplitude weakly nonlinear LSs computed near the Turing bifurcation persist as they enter the nonlinear regime, leading to complex bifurcation structures known as {\it homoclinic snaking} \cite{woods_heteroclinic_1999, burke_snakes_2007, knobloch_homoclinic_2005}. The region of existence of these states, referred to hereafter as the {\it localization region}, is illustrated in the $(\alpha,\beta)$ phase diagram shown in Fig.~\ref{fig_phase_dia_1} for $\epsilon=1$ and $\delta=0.7$.

Due to the symmetry of our model, there are two localization regions: one for positive values of $\beta$, corresponding to holes, and a symmetrically opposed region for $\beta<0$, corresponding to bumps or peaks. Within these regions, and for the parameter range we have studied, three main sectors can be distinguished, each associated with different types of LSs and corresponding bifurcation structures. The localization region is divided into the following three sectors (see Fig.~\ref{fig_phase_dia_1}):
\begin{itemize} \item [I:] {\it Standard homoclinic snaking region} \item [II:] {\it Transition region} \item [III:] {\it Collapsed homoclinic snaking region} \end{itemize}

In the following, we will analyze the complexity of these scenarios by classifying the different structures and determining their stability. This structure is generic in systems exhibiting multistability between two uniform states and a spatially periodic pattern, as demonstrated in Refs.~\cite{parra-rivas_organization_2023,akakpo_implications_2024}.

	\begin{figure*}[!t]
		\centering
		\includegraphics[scale=0.94]{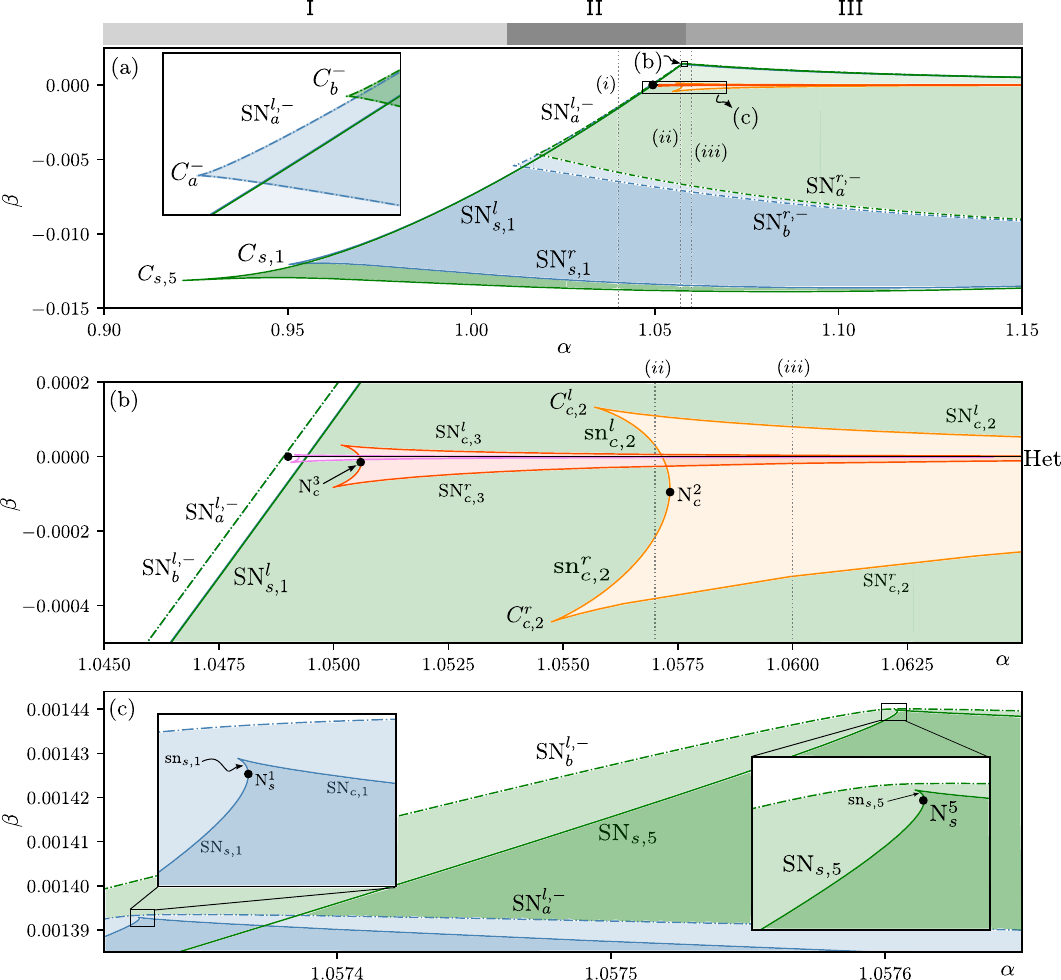} 
		\includegraphics[scale=0.94]{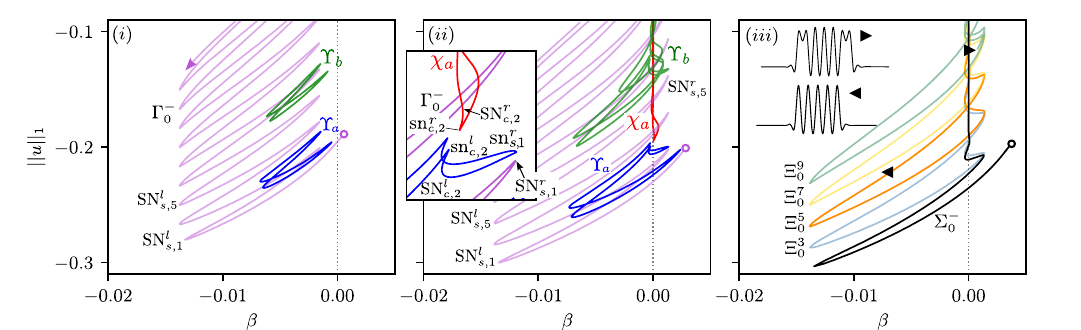} 
        \caption{\textbf{Transition region between standard and collapsed homoclinic snaking.} (a) Close-up view of the diagram shown in Fig.~\ref{fig_phase_dia_1} around the existence region of the peak LSs. (b) Close-up view of (a) focusing on the necking bifurcations of the standard snaking. (c) Close-up view of the necking bifurcations associated with the collapsed-snaking-like LSs. Panels (i)-(iii) show one-parameter bifurcation diagrams illustrating the transition for $\alpha=1.04$, $1.057$, and $1.06$, respectively.}
		\label{fig_transition}
	\end{figure*}

			\begin{figure*}[!t]
		\centering
		\includegraphics[scale=1]{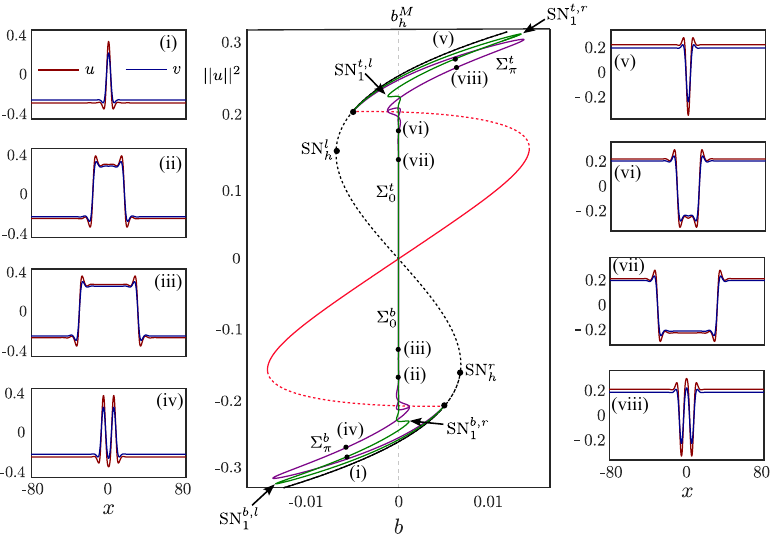}
        \caption{{\bf Collapsed homoclinic snaking.} This bifurcation diagram corresponds to the parameters $(\epsilon,\delta,\alpha)=(1,0.7,1.07)$. It displays the HSSs $U_h^{b,m,t}$ (black), the periodic Turing pattern (red) arising from TI$_{1,2}$, its saddle-node bifurcations SN$P^{l,r}$, and the two families of LSs, $\Sigma$ and $\Sigma'$, which undergo collapsed snaking.
Each group is formed by the families $\Gamma{0,\pi}^{b,t}$, where the superscript $b$ ($t$) corresponds to bottom (top) LSs. For each snaking, a close-up view is provided, with the spectral stability of the states indicated using solid (dashed) lines for stable (unstable) states. Sample LS profiles along these families are shown in panels (i)-(iv) for the bump states and panels (v)-(viii) for the hole states. The vertical dashed line $b_M^h$, at $b=0$, marks the uniform Maxwell point between the HSSs $U_h^{b,t}$.}
		\label{fig_collapsed}         
	\end{figure*}

\subsection{Standard homoclinic snaking}

In sector I, LSs organize into a bifurcation structure known as standard homoclinic snaking \cite{woods_heteroclinic_1999,burke_snakes_2007}. An example of this configuration is depicted in Fig.\ref{fig_standard} for $\alpha=1$, showing the modification of the $L_1$-norm of $u$:
	\begin{equation}
||u||_{L_1}\equiv\frac{1}{L}\int_{-L/2}^{L/2}u(x)dx,
	\end{equation}
     as a function of $\beta$. Here, LSs form due to the pinning of patterned fronts connecting the uniform state with a subcritical periodic pattern. For the value of $\alpha$ considered here, this pattern emerges subcritically and unstably from TI$^\pm$ and becomes stable at the saddle-node bifurcations SN$_P^{l,r}$. Two examples of these pattern profiles are depicted in Figs.\ref{fig_standard}(i) and \ref{fig_standard}(v). This configuration yields two bistable regions (gray-shaded areas) where LS locking occurs, and two distinct snaking structures emerge from TI$^\pm$: $\Gamma^{+}$ at the top for $\beta>0$, and $\Gamma^{-}$ at the bottom for $\beta<0$. Examples of LSs along these diagrams are shown in Figs.~\ref{fig_standard}(ii)-(iv) and \ref{fig_standard}(vi)-(viii). These states resemble a slug of the spatially periodic pattern embedded in a uniform surrounding.

For each snaking structure, there are two families of homoclinic snaking curves, labeled $\Gamma^{\pm}_{0,\pi}$. The curves with the subscript $0$ (i.e., $\Gamma^{\pm}_0$) are associated with LSs consisting of an odd number of pattern rolls or peaks. Here, we show two examples containing three peaks [see Fig.\ref{fig_standard}(ii)] and three holes [Fig.\ref{fig_standard}(vii)]. All these nonlinear states are homotopically connected to the small-amplitude asymptotic states computed in Section~\ref{sec:4} when $\varphi=0$ [see Eq.~(\ref{loc_states})]. The curves $\Gamma^{\pm}_\pi$, homotopically connected to the weakly nonlinear states with $\varphi=\pi$, correspond to LSs with an even number of pattern rolls [see profiles depicted in Figs.~\ref{fig_standard}(ii),(iv),(vi), and (viii)].

Along these snaking diagrams, LSs gain or lose stability when crossing the left and right folds, corresponding to saddle-node bifurcations SN$^{l,r}_{i-}$ at the bottom and SN$^{l,r}_{i+}$ at the top, with $i$ denoting the number of pattern rolls in the LS profiles. Stability is indicated using solid (dashed) lines for stable (unstable) states. These stability changes arise from the complex heteroclinic tangle process underlying the formation of these states \cite{woods_heteroclinic_1999,coullet_stable_2000, gomila_bifurcation_2007}. Throughout the pinning interval, which approximately spans the entire region of bistability between the patterned and uniform state, 
the system exhibits LS multistability.

This structure persists throughout region I and begins to disappear in a complex process involving a sequence of codimension-2 necking bifurcations \cite{parra-rivas_organization_2023}, which we will describe in the following sections.

\subsection{Transition region}
We refer to the parameter region where there is an overlap of elements corresponding to the standard and collapsed homoclinic snaking \cite{parra-rivas_organization_2023} as the transition sector. Its definition is approximate rather than rigorous, and we label this as Sector II. To explain, we focus on Fig.~\ref{fig_transition}(a), which shows this sector in detail for bump states, i.e., $\beta < 0$. The case for $\beta > 0$ is mirror-symmetric with respect to the axis $\beta = 0$.

On the left, this region is bounded by a sequence of cusp bifurcations $C_{a,b}^-$ [see the close-up view in Fig.~\ref{fig_transition}(a)], from which new isolas arise. These isolas coexist with $\Gamma_{0,\pi}^{-}$. Two examples of these isolas, along with some of their associated states, are illustrated in Fig.~\ref{fig_transition}(i) for $\alpha = 1.04$. We refer to them as $\Upsilon_a^-$ and $\Upsilon_b^-$. The extent of these isolas is defined by the saddle-node bifurcations SN$_a^{l,r}$ and SN$_b^{l,r}$, which are depicted in Fig.~\ref{fig_transition}(a) using point-dashed lines.

Within this region, additional bifurcations, essential to this transition, occur in very narrow parameter ranges highlighted by square boxes [see the top of Fig.~\ref{fig_transition}(a)]. Enlarged views of these areas are shown in Figs.~\ref{fig_transition}(b) and \ref{fig_transition}(c).

As $\alpha$ increases, SN$_{a,b}^{l}$ and SN$_{a,b}^{r}$ diverge, causing $\Upsilon_a^-$ and $\Upsilon_b^-$ to expand. Simultaneously, other isolas ($\chi$) associated with the locking of uniform fronts begin to emerge from the point $(\alpha,\beta)=(\alpha_X,\beta_X)\approx (1,??)$. This codimension-two point marks the initiation of the locking process between uniform fronts of different polarity. An example of a $\chi$-isola is shown in Fig.~\ref{fig_transition}(ii) for $\alpha=1.057$. The origin of $\chi$ isolas is tied to the occurrence of cusp bifurcations ($C_{c,i}^{l,r}$) located in the fork-like regions illustrated in Fig.~\ref{fig_transition}(b).

For instance, from $C_{c,2}^{l}$, the bifurcations SN$_{c,2}^l$ and sn$_{c,2}^l$ arise, and a similar process occurs for $C_{c,2}^{r}$. These bifurcations are depicted in the close-up view of Fig.~\ref{fig_transition}(ii). As $\alpha$ increases further, sn$_{c,2}^l$ and sn$_{c,2}^r$ merge at N$_c^2$, a codimension-two bifurcation known as a necking bifurcation \cite{parra-rivas_organization_2023}. This results in the merging of $\Upsilon_a$ and $\chi_a$, forming a new type of isola, $\Xi_0^1$, not shown here. Similar processes involving other $\chi$-like isolas originating from $C_{c,3}^{l,r}$ and merging at N$_c^3$ are discussed in Ref.~\cite{parra-rivas_organization_2023}.

Further increasing $\alpha$, SN$_{s,1}^r$ in $\Gamma_0^-$ and sn$_{s,1}$ in $\Xi_0^1$ collide in another necking bifurcation, N$_s^1$, resulting in the formation of a new structure $\Sigma_0^-$, referred to as collapsed snaking \cite{knobloch_homoclinic_2005}. The location of this bifurcation in the $(\alpha,\beta)$ parameter space is depicted in Fig.~\ref{fig_transition}(c). The organization of this new bifurcation structure will be detailed in the next section.

Similarly, other isolas undergo merging processes that destroy $\Gamma_0^-$. For completeness, we illustrate the example of $\Upsilon_b$ [see Figs.~\ref{fig_transition}(i)-(ii)]. Eventually, this isola merges with the remnants of the standard homoclinic snaking at N$_s^5$, forming $\Xi_0^5$. Along this isola, states of various types, including hybrid states, emerge, as shown in Fig.~\ref{fig_transition}(iii). Other isolas, such as $\Xi_0^3$, $\Xi_0^7$, and $\Xi_0^9$, form through similar processes and persist at larger values of $\alpha$, coexisting with collapsed homoclinic snaking.

Sector II expands until the final necking bifurcation, N$_s^i$, and the complete destruction of the standard homoclinic snaking.

\subsection{Collapsed homoclinic snaking}
The collapsed snaking formed in the transition sector [see Fig.~\ref{fig_transition}(iii)] persists throughout all of sector III. In this region, a new type of LS arises from the interaction of uniform fronts \cite{coullet_nature_1987,coullet_localized_2002,knobloch_homoclinic_2005}. An example of this bifurcation structure is shown in Fig.~\ref{fig_collapsed} for $\alpha=1.07$. Similar to the standard homoclinic snaking case, a pair of snaking curves $\Sigma^\pm_0$ emerges from the Turing bifurcation points TI$^\pm$, connecting with the small-amplitude state (\ref{loc_states}) with $\varphi=0$. Likewise, $\Sigma^\pm_\pi$, associated with $\varphi=\pi$, also appears.

\begin{figure*}[!t]
	\centering
	\includegraphics[scale=0.95]{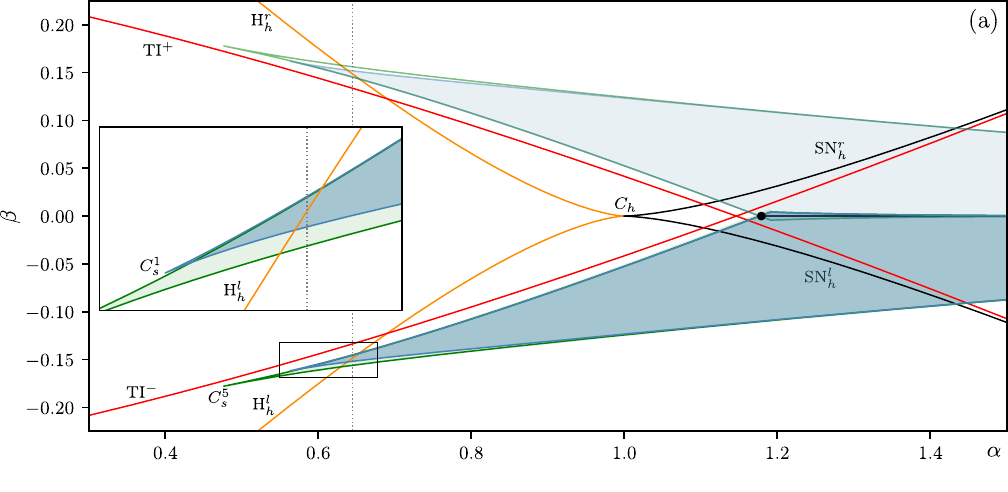} 
	\includegraphics[scale=1]{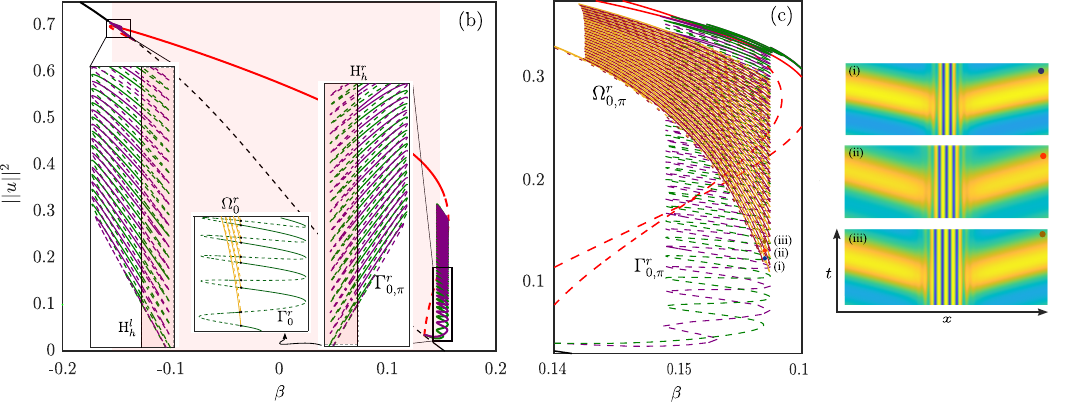} 
    \caption{{\bf Bifurcation structure of the Turing-Hopf localized states.} (a) Phase diagram in the $(\alpha,\beta)$-parameter space for $\epsilon=1$ and $\delta=0.4$, showing the main bifurcation curves of the system as in Fig.~\ref{fig_phase_dia_1}: TI (red), SN$_h^{l,r}$ (black), H$_h$ (orange), and SN$_{1}^{l,r;t,b}$ for the localization regions. We also mark the codimension-2 bifurcations: $C_h$, $C_{b,t}$, TI$^d_{1,2}$, and the codimension-2 heteroclinic point Het$_p$. The close-up view highlights the new oscillatory region associated with the standard homoclinic snaking-related LSs. (b) Bifurcation snaking diagram for Turing-Hopf LSs when $(\epsilon,\delta,a)=(1,0.4,0.645)$. The green and purple dashed curves correspond to the standard homoclinic snaking associated with time-independent unstable LSs, i.e., $\Gamma_{0,\pi}$. (c) Same as (b), but for $(\epsilon,\delta,a)=(1,0.4,0.628)$. The brown and orange curves represent the Turing-Hopf LSs, with solid (dashed) lines denoting stable (unstable) oscillatory states. Examples of these oscillatory states are illustrated in panels (i)-(iii).}
	\label{fig_phase_dia_2}
\end{figure*}	

Let us now focus on $\Sigma_0^-$. The small-amplitude unstable LS centered at $x=0$ increases its amplitude as $\beta$ decreases and stabilizes at the first left fold, SN$_{c,1}^{l}$. An example of this single-peak state is shown in Fig.~\ref{fig_collapsed}(i). As $\beta$ increases, this state continues to grow in amplitude and becomes unstable at SN$_{c,1}^{r}$. Beyond this fold, a broadening process occurs as we move up the bifurcation diagram: at each right fold, a new spatial oscillation or dip nucleates, leading to increasingly wider LSs \cite{parra-rivas_dark_2016}. Examples of these types of states along the diagram are depicted in Fig.~\ref{fig_collapsed}(ii) and Fig.~\ref{fig_collapsed}(iii), where the underlying uniform fronts forming these states can be identified [see Fig.~\ref{fig_collapsed}(iii)]. This process is associated with damped oscillations around the uniform Maxwell point $\beta^M_u$ and specifically around the heteroclinic (Het) connection, which results from an interaction and locking process \cite{parra-rivas_origin_2021,parra-rivas_organization_2023}. The exponentially decreasing amplitude of these oscillations gives rise to the term "collapsed snaking" \cite{knobloch_homoclinic_2005,burke_classification_2008,parra-rivas_dark_2016}.

Similarly, the small-amplitude states arising from TI$^+$ develop into the collapsed snaking $\Sigma_0^+$. In this case, LSs appear as holes, as illustrated in Fig.~\ref{fig_collapsed}(v)-(viii).

The spectral stability analysis along this curve reveals that LSs are temporally stable between SN$_{c,i}^{l}$ and SN$_{c,i}^{r}$ for the same $i$, and unstable between SN$^{r}_{c,i}$ and SN$^{l}_{c,i+1}$, as indicated by solid (stable) and dashed (unstable) lines.

In an infinite system, the dip nucleation process continues indefinitely, and $\Sigma_0^\pm$ never connect to one another. However, in the finite domain analyzed here, this process ends when the two front waves reach the boundaries of the domain, connecting the two snaking structures.

As $\alpha$ increases, the collapsed snaking becomes more prominent. This trend can be observed in Fig.~\ref{fig_transition}(b), where SN$_{c,i}^l$ and SN$_{c,i}^r$ approach one another and eventually collide sequentially. When this happens, LSs with $i$ bumps disappear, and the collapsed snaking reduces to a monotonic vertical line at Het. This entire process is related to the occurrence of a BD transition, where the oscillatory tails of the uniform fronts vanish, eliminating the possibility of front locking and LS formation. For the parameter set used here, this transition occurs at $\alpha_{\rm BD}\approx 1.4286$. 

A similar fate is observed for the remnants of the homoclinic snaking and hybrid states organized in the $\Xi$-isolas, whose region of existence shrinks as $\alpha$ increases and eventually vanishes.


    \section{Oscillatory dynamics involving localized states}\label{sec:5}
As we have seen previously, by varying either $\delta$ or $\epsilon$, we can modify the relative position of the Hopf and Turing bifurcations, leading to new dynamical behaviors that were absent in the previous configurations. In this section, we explore and characterize two new dynamical regimes that give rise to oscillatory LSs of different natures. In one regime, the background field undergoes periodic oscillations in time while the LS remains static. In the second scenario, the background remains stable while the LS oscillates. 
	
\subsection{Turing-Hopf localized patterns}
Let us first consider the first type of oscillatory states. Figure~\ref{fig_phase_dia_2} illustrates the modification of the $(\alpha, \beta)$ phase diagram shown in Fig.~\ref{fig_phase_dia_1} when $\delta$ is reduced to $\delta=0.4$. As $\epsilon$ is kept constant at $\epsilon=1$, the position of the Hopf line remains unchanged. However, the reduction in $\delta$ causes a leftward shift in the Turing instability and the entire localization region. As a result, the Hopf line intersects the localization region, creating a new dynamical sector characterized by oscillatory behavior (see sector X in Fig.~\ref{fig_phase_dia_2}(a)). Figure~\ref{fig_phase_dia_2}(b) provides an example of how the standard homoclinic snaking is modified in this region. To better illustrate this structure, we use the $L_2$-norm:
	\begin{equation}
||u||_{L_2}^2=\frac{1}{L}\int_{-L/2}^{L/2}u(x)^2dx.
	\end{equation}
The two vertical solid lines in both close-up views indicate the positions of the uniform Hopf bifurcations H$_h^{l,r}$. In the region between these two lines (highlighted in pink), the uniform state becomes unstable. Additionally, this instability affects all the $\Gamma_{0,\pi}$ branches, destabilizing portions of them.

From each Hopf bifurcation on the stable branches of the homoclinic snaking, Turing-Hopf LSs emerge supercritically (see Fig.~\ref{fig_phase_dia_2}(b)). These oscillatory states consist of static localized patterns embedded within an oscillatory background. Examples of these states are shown in Fig.~\ref{fig_phase_dia_2}(i)-(iii). As the oscillation amplitude grows, each of these states increases its norm while moving further from the Hopf bifurcations H$^h_i$ (see solution branches $\Omega_0^r$).

Focusing on the single-peak Turing-Hopf state depicted in Fig.~\ref{fig_phase_dia_2}(b), this state emerges stably from the single-peak branch of $\Gamma_0^b$ and eventually undergoes a fold of cycles (FC), where its stability changes (not shown here). The stability is computed through a Floquet analysis performed using the path-continuation software \texttt{pde2path} \cite{uecker_numerical_2021}. This branch of unstable oscillations folds back, decreasing its norm while approaching H$_h^{1,2}$ until it reconnects with $\Gamma_0^-$ on the unstable single-peak branch. This branching behavior has also been observed in other reaction-diffusion systems, such as the Gilad-Meron model for plant ecology \cite{al_saadi_transitions_2023}, and has been explained in detail in a predator-prey model \cite{al_saadi_time-dependent_2024}. This process continues similarly as one moves up the snaking structure, leading to the emergence of additional Turing-Hopf LSs.

	\begin{figure*}[!t]
		\centering  
		\includegraphics[scale=1]{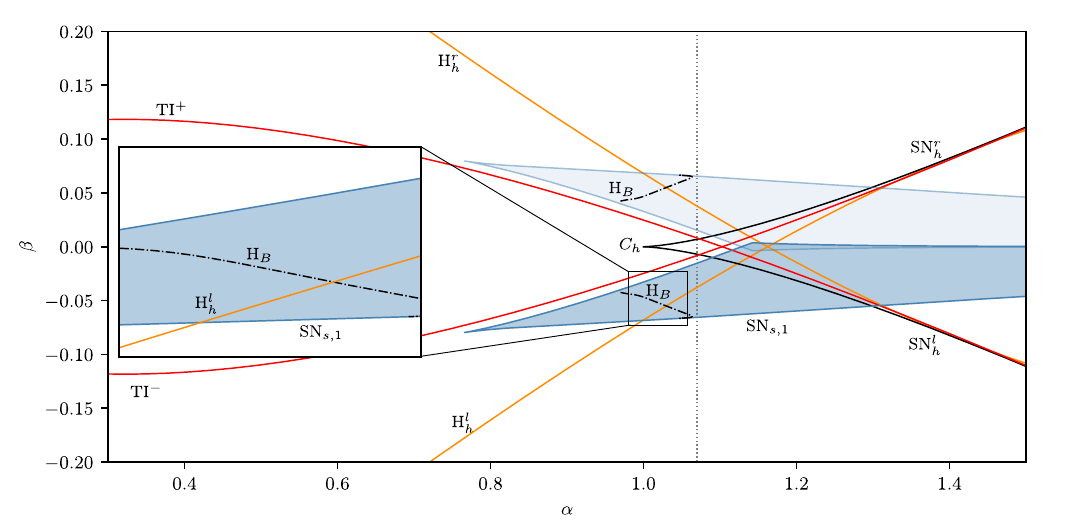}
				\includegraphics[scale=1]{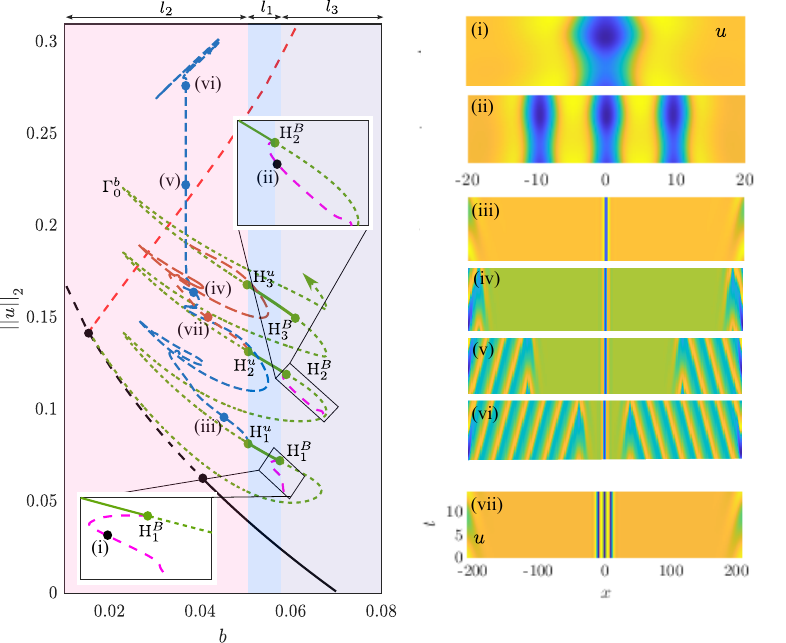}
		\caption{\textbf{Bifurcation structure of unstable breathers.} (a) Phase diagram in the $(\alpha,\beta)$-parameter space for $\epsilon=0.5$ and $\delta=0.7$, showing the main bifurcation curves of the system, as in Fig.~\ref{fig_phase_dia_1}: TI (red), SN$h^{l,r}$ (black), H$h$ (orange), and SN${1}^{l,r;t,b}$ for the localization regions. We also mark the codim-2 bifurcations: $C_h$, $C{b,t}$, TI$^d_{1,2}$, and the codim-2 heteroclinic point Het$_p$. The close-up view highlights the new oscillatory region for the standard homoclinic snaking-related LSs. The vertical line corresponds to the bifurcation diagram shown in (b). (b) Bifurcation diagram for oscillatory states when $(\epsilon,\delta,a)=(0.5,0.7,1.07)$. The standard homoclinic snaking $\Gamma^b_0$ is partially stable in the blue-shaded area. H$_i^B$ denotes the Hopf bifurcation leading to breathers, while H$_i^u$ corresponds to the uniform Hopf instability. Panels (i) and (ii) show breather-like states, with one oscillation period depicted. Panels (iii)-(vi) illustrate the variation of Turing-Hopf LSs with a single peak along $\Sigma^O$. Panel (viii) presents another example of a Turing-Hopf state with three peaks.}
		\label{fig_phase_dia_3}
	\end{figure*}

Decreasing $\alpha$ and crossing the uniform Hopf line renders the standard homoclinic snaking completely unstable. An example of this situation is depicted in Fig.~\ref{fig_phase_dia_2}(c), corresponding to the vertical line shown in Fig.~\ref{fig_phase_dia_2}(a) for $\alpha=0.628$. Here, the previously disconnected stable and unstable pairs of Turing-Hopf LS branches (see close-up in Fig.~\ref{fig_phase_dia_2}(c)) reorganize, yielding the snaking curves $\Omega_{0,\pi}^r$ illustrated in Fig.~\ref{fig_phase_dia_2}(c) (see brown and orange curves). Further along the diagram, the localized oscillations exhibit a morphology similar to their time-independent counterparts: two families of curves emerge, one associated with Turing-Hopf states with an odd number of peaks and the other with an even number. Examples of these states along $\Omega_{0,\pi}^r$ are shown in Fig.~\ref{fig_phase_dia_2}(i)-(iii).

The homoclinic snaking of Turing-Hopf LSs has been analyzed in detail in the context of the Brusselator model \cite{tzou_homoclinic_2013} and more recently in a predator-prey model \cite{al_saadi_time-dependent_2024}. Interested readers are encouraged to consult these works for further details.

\subsection{Unstable breathers}
The previously described Turing-Hopf LSs are not the only oscillatory states present in the FHN model. By exploring the entire parameter space of the system, we have identified another type of oscillatory dynamics where the LS oscillates periodically, while the background field remains unaffected. These oscillations are commonly referred to as {\it breathers} or {\it oscillons} in the nonlinear dynamics literature \cite{aranson_patterns_2006, parra-rivas_parametric_2020-1}.

In the FHN equations, these states appear when $\delta=0.7$ and $\epsilon$ is reduced to $\epsilon=0.5$. Under these conditions, the phase diagram depicted in Fig.~\ref{fig_phase_dia_1} transforms into the one shown in Fig.~\ref{fig_phase_dia_3}(a). Here, the localized region intersects the uniform Hopf line, similar to the situation in Fig.~\ref{fig_phase_dia_2}, leading to the emergence of Turing-Hopf LSs. Additionally, LSs encounter another Hopf instability, H$^B_i$, which gives rise to breather states. Note that this instability is unrelated to the uniform state Hopf bifurcation. These two Hopf lines divide the localization region into four main regions: $l_1$, where LSs are time-independent states; $l_2$, where Turing-Hopf states exist; $l_3$, where only breathers appear; and $l_4$, where mixed oscillatory behaviors emerge.

Let us take a closer look at this mixed oscillatory regime. Figure~\ref{fig_phase_dia_3}(c) shows the bifurcation diagram corresponding to a slice of Fig.~\ref{fig_phase_dia_3} at $\alpha=1.07$. Here, the standard homoclinic snaking $\Gamma_0^-$ is almost entirely unstable due to oscillatory instabilities. Only within the blue-shaded area, corresponding to $l_1$, are odd localized patterns stable. 

To the right of this stable region, LSs undergo Hopf bifurcations (H$^B_i$, $i=1,2,3$), associated with breather-like behavior (see $l_3$). From the first of these bifurcations, H$^B_1$, a breather state corresponding to an oscillatory single-peak state emerges, though it is always unstable [see a single period of this oscillation in Fig.~\ref{fig_phase_dia_3}(i)]. Similar behavior is observed near H$^B_2$ and H$^B_3$, where unstable breathers with 3 and 5 peaks, respectively, emerge. An example of the 3-peak LS arising from H$^B_2$ is shown in Fig.~\ref{fig_phase_dia_3}(ii). However, across the entire parameter regime explored here, breathers are consistently found to be unstable.

To the left of $l_1$, in region $l_2$, Turing-Hopf LSs emerge supercritically from H$^u_i$ and follow intricate bifurcation curves such as $\Sigma^O_{1,2}$ in Fig.~\ref{fig_phase_dia_3}(b). The Turing-Hopf states along $\Sigma^O_{1}$ exhibit a collapsed-like snaking behavior, leading to modifications illustrated in Figs.~\ref{fig_phase_dia_3}(iii)-(vi). The oscillations of the uniform background initially appear at the domain boundaries and progressively approach the central peaks as one follows the $\Sigma^O_{1}$ diagram, eventually rendering all uniform states unstable [see sequence (iii)-(vi)]. Similar behavior is observed for the 3-peak Turing-Hopf LS emerging from H$^u_2$, which follows the bifurcation structure $\Sigma^O_2$. All these oscillatory states appear to be completely unstable as well.

	\begin{figure*}[!t]
		\centering
		\includegraphics[scale=1.1]{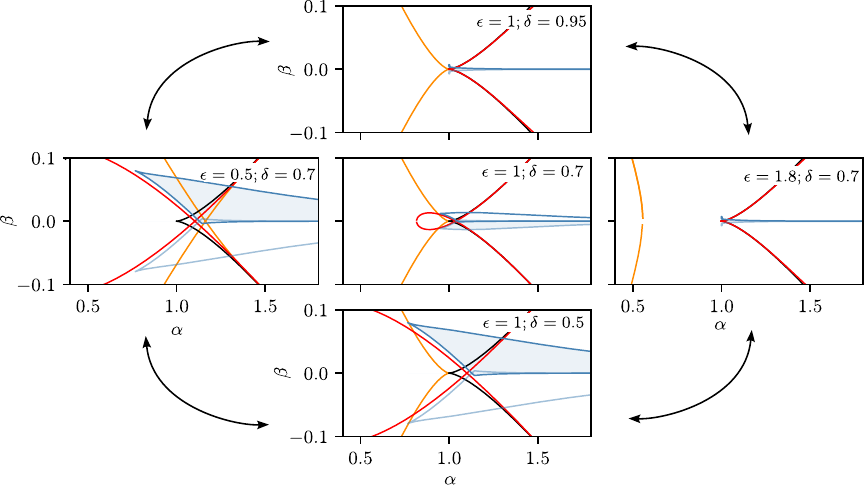}
        \caption{{\bf Modification of the $(\alpha,\beta)$-phase diagrams with $\epsilon$ and $\delta$.} The red curves represent the TI$^\pm$; the orange curve corresponds to the uniform Hopf bifurcation; the black lines indicate SN$_h^{l,r}$, and the blue regions denote the localization zones for holes and bumps.}
		\label{mapfinal}
	\end{figure*}

\section{Modification of the dynamical regions with $\epsilon$ and $\delta$}\label{sec:6}

In the previous sections, we investigated the formation and dynamics of LSs in the parameter plane $(\alpha,\beta)$ for some representative values of $\epsilon$ and $\delta$. However, we do not yet have a complete picture of how these dynamics change in a four-dimensional parameter space, i.e., when incorporating the effects of both $\epsilon$ and $\delta$. Our aim in this section is to clarify this point. Figure~\ref{mapfinal} illustrates such modifications. In the center column, we fix $\epsilon$ to $1$ and allow $\delta$ to vary. Conversely, in the central row, $\delta$ is fixed while $\epsilon$ changes. For simplicity, oscillatory instabilities of the LSs are not included in this analysis. 

Let us start in the center of this diagram, where $(\epsilon,\delta)=(1,0.7)$. This is the same diagram as the one depicted in Fig.~\ref{fig_phase_dia_1}. The shaded blue areas represent the localization regions of the system, while the other bifurcation lines are TI$^\pm$ in red, uniform Hopf H$_h^\pm$ in orange, and SN$_h^{l,r}$ in black, which here is partially overlapped with TI$^\pm$. As we learned in previous sections, the pair $(\alpha,\beta)$ triggers the transition from a single uniform solution to three. This transition is not affected by the modification of $\epsilon$ and $\delta$. Thus, the loci of SN$_h^{l,r}$ remain invariant under these parameters (compare all diagrams). Furthermore, the position of H$_h^\pm$ is not affected by $\delta$ but depends only on $\epsilon$ and $\alpha$. Therefore, the loci of H$_h^\pm$ remain invariant when moving vertically along the diagrams. However, when moving horizontally across the diagrams, both H$_u^\pm$ and TI$^\pm$ are modified.

Now, let us examine what happens to the localization regions as $\delta$ and $\epsilon$ vary. Decreasing the diffusion parameter $\delta$ (see the central column in Fig.~\ref{mapfinal}) enhances the emergence of LSs and increases their region of existence and stability. A similar effect occurs when $\delta$ is fixed, and $\epsilon$ is decreased: the localization region expands. In contrast, increasing $\delta$, $\epsilon$, or both has a very negative effect on the formation of LSs.

\section{Discussion and conclusions}\label{sec:7}
In this work, we have presented a detailed study regarding the formation and bifurcation structure of LSs in the 1D FHN model. Despite the long history of this model, few studies reported on LSs \cite{cebrian-lacasa_six_2024,frohoff2023stationary}, and a detailed study focused on LSs was lacking in the existing literature \cite{cebrian-lacasa_six_2024}.
Here, we have used bifurcation analysis to unveil the origin and bifurcation structure of LSs formed when there is multistability between different uniform and extended patterned solutions \cite{parra-rivas_organization_2023}.

This model undergoes a transition between uniform-pattern and uniform-uniform regions in the parameter space. In the uniform-pattern configuration, LSs emerging from subcritical Turing instabilities undergo standard homoclinic snaking characterized by two bifurcation curves, $\Gamma_{0,\pi}$ (see Sec.~\ref{sec:4}), which oscillate back and forth within the pinning region. In contrast, in the uniform-uniform configuration, LSs exhibit collapsed snaking. Here, the morphology of the bifurcation curves, $\Sigma_{0,\pi}$, corresponds to damped oscillations around the uniform Maxwell point of the system (see Sec.~\ref{sec:4}). In both cases, the bifurcation structure is closely related to the front-locking mechanism underlying LS formation. Varying the control parameters reveals a smooth transition between these two scenarios, mediated by a cascade of codimension-two necking bifurcations, similar to those discovered in other pattern-forming systems such as the Swift-Hohenberg equation \cite{parra-rivas_organization_2023} and the Lugiato-Lefever equation \cite{akakpo_implications_2024}. The agreement between these and previous results confirms that the transition scenario described here is generic in systems displaying uniform-pattern-uniform tristable configurations.

Beyond these transitions, this system exhibits other interesting dynamical behaviors, including two types of time-dependent oscillatory LSs. In one case, the oscillatory states consist of steady LSs embedded in an oscillatory background field. We refer to these as Turing-Hopf states. Such states have been predicted in other systems, including the Swift-Hohenberg equation \cite{tzou_homoclinic_2013} and other reaction-diffusion systems \cite{borckmans_localized_1995,de_wit_spatiotemporal_1996}, as well as the Gilad-Meron model for semi-arid plant ecology \cite{al_saadi_transitions_2023,al_saadi_time-dependent_2024}. In the second case, the background remains static, and the LS itself oscillates in amplitude. These are the so-called breathers in the optics literature or oscillons. Despite their intriguing bifurcation structure, these states remain unstable in the parameter regime analyzed.

A natural extension of this work is to include configurations in higher dimensions, which would make the present study more realistic. One fundamental question to address regards the existence of 2D and 3D generalizations of the states we have found here. If such states exist, do they preserve the same bifurcation structure? Even if they do, LSs may undergo curvature-related instabilities absent in the 1D case. To carry out this future study, we will first focus on radially symmetric configurations. In this context, a very useful approach is to use the system's dimensionality as a control parameter \cite{mccalla_snaking_2010}. This method allows homotopical connections between states of different dimensions, enabling predictions of higher-dimensional states based on the 1D results presented here.

Another promising extension of this work is the inclusion of mass conservation to study how the bifurcation structure of spatially localized structures is altered in the presence of such a constraint. The study of active matter systems has advanced significantly in recent years, where energy input drives self-propelled motion, leading to non-reciprocal field coupling and inherently non-variational dynamics
\cite{Fruchart2021,Dinelli2023,PhysRevX.10.041009,PhysRevX.10.041036,PhysRevX.14.021014,PhysRevLett.131.107201,frohoff2023stationary}. These characteristics have inspired the development of the non-reciprocally coupled Cahn-Hilliard equation to model pattern formation in biological non-equilibrium systems \cite{PhysRevX.10.041009,PhysRevX.10.041036,PhysRevX.14.021014,PhysRevLett.131.107201,frohoff2023stationary}. Interestingly, this equation is mathematically analogous to a mass-conserved FHN model. Even the standard FHN model with non-conserved dynamics studied here is non-variational, highlighting its relevance to active systems. Extending our current analysis to incorporate mass conservation could provide new insights into the dynamics and bifurcation structures of LSs in active matter systems.

In conclusion, we hope that this detailed bifurcation analysis, which identifies parameter regimes where LSs exist, will serve as a valuable resource for experimentalists seeking such phenomena in biological systems modeled by the FHN equations \cite{cebrian-lacasa_six_2024}. 

We find that stronger time-scale separation and greater differences in the diffusion coefficients between fields expand the parameter regions supporting the existence of LSs (Fig.~\ref{mapfinal}).
While numerous studies have explored extended patterns, traveling waves, and pulses in the FHN model \cite{BetaKrusereviewwaves,64, ditalia2022,189,173,nolet2020,70,148,149, 175}, research on localized patterns in the FHN model and its biological applications remains relatively sparse.

Recent theoretical work has described mechanochemical localized states, where gradients in active stress induce flows that advect chemically regulated assemblies of active matter \cite{Barberi_PRL_2023,Barberi_PRE_2024}. Establishing connections between these theoretical predictions and corresponding experimental observations could provide a deeper understanding of spatially localized phenomena in diverse biological contexts.

\section*{Acknowledgements}
The work is supported by Internal funds KU Leuven (C14/23/130, L.G.), and a junior research grant from the Research Foundation – Flanders (G074321N, L.G.). 

\section*{Data and code availability}
?

\appendix	
	
\section{Linear stability analysis of the uniform state and the dispersion relation}\label{append1}
Equation~(\ref{model}) can be written compactly in matrix form as: 
\begin{equation}
	\left[\begin{array}{c}
		\partial_tu \\ \partial_tv
	\end{array}
	\right]=\left(\mathcal{L}+\mathcal{N}\right) \left[\begin{array}{c}
		u\\ v
	\end{array}
	\right]+\mathcal{Y}
,
\end{equation}
where the linear operator is defined as:
\begin{equation}
	\mathcal{L}=\mathcal{A}\partial_x^2+\mathcal{B},
\end{equation}
with
$$\mathcal{A}\equiv\left[\begin{array}{cc}
	\delta^2 & 0  \\ 0 & 1
\end{array}\right],\qquad\mathcal{B}\equiv\left[\begin{array}{cc}
1 & -1  \\ \varepsilon & -\varepsilon\alpha
\end{array}\right].$$
The nonlinear operator and constant terms are
\begin{equation}
	\mathcal{N}\equiv -u^2\mathcal{M}\equiv- u^2 \left[\begin{array}{cc}
		1 & 0  \\ 0 & 0
	\end{array}\right], \quad 	\mathcal{Y}\equiv \left[\begin{array}{c}
		0\\ -\beta
	\end{array}
	\right].
\end{equation}

To perform the linear stability analysis of the uniform state $(U_h,V_h)$ we introduce a perturbation: 
$$	\left[\begin{array}{c}
	u(x,t)\\v(x,t)
\end{array}\right]=	\underbrace{\left[\begin{array}{c}
U_h\\V_h
\end{array}\right]}_{\displaystyle\vec{Q}}+	\epsilon\underbrace{\left[\begin{array}{c}
\phi(x,t)\\\psi(x,t)
\end{array}\right]}_{\displaystyle\vec{q}(x,t)}.$$
Expanding the nonlinear operator as
$$\mathcal{N}=\mathcal{N}+\epsilon \mathcal{N}_1+\mathcal{O}(\epsilon^2),$$
with 
$$\mathcal{N}_0\equiv -U_h^2\mathcal{M},\quad\mathcal{N}_1\equiv-2U_h\phi\mathcal{M},$$
the linear operator is $\mathcal{O}(\epsilon^0)$, i.e. $\mathcal{L}=\mathcal{L}_0$.

Inserting these expansions in Eq.~(1) we obtain the following hierarchy of equations:
$$
\begin{array}{ll}
	\mathcal{O}(\epsilon^0):& 0=\left(\mathcal{L}_0+\mathcal{N}_0\right)\vec{Q}_0+\vec{\mathcal{Y}}_0\\\\
	\mathcal{O}(\epsilon^1):& \partial_t\vec{q}=\left(\mathcal{L}_0+\mathcal{N}_0\right)\vec{q}+\mathcal{N}_1\vec{Q}.\\
\end{array}
$$
The equation at $\mathcal{O}(\epsilon^0)$ defines the uniform state solution (HHS) given by Eq.~(\ref{hom_sol}). Considering the equality $\phi\mathcal{M}\vec{Q}=U_h\mathcal{M}\vec{q}$, we have that $\mathcal{N}_1\vec{Q}=-2U^2_h\mathcal{M}\vec{q}$, and the equation at  $\mathcal{O}(\epsilon^1)$ becomes
\begin{equation}\label{eq.L}
\partial_t\vec{q}=L\vec{q},
\end{equation} 
where the linear operator reads
\begin{equation}
L\equiv\mathcal{L}_0+3\mathcal{N}_0= \left[\begin{array}{cc}
	\delta^2\partial_x^2+1-3U_h^2 & -1 \\ \varepsilon & \partial_x^2-\varepsilon\alpha
\end{array}\right].
\end{equation}
If we now consider perturbations of the form 
$\vec{q}=\vec{\xi}e^{\sigma t+ikx}+c.c.$, Eq.~(\ref{eq.L}) becomes 
$$\left(L^{(1)}-\sigma\mathbf{I}_{2\times2}\right)\vec{\xi}=\vec{0},$$
where 
$$
L^{(1)}\equiv\left[\begin{array}{cc}
	L^{(1)}_{11} & -1 \\ \varepsilon & L^{(1)}_{22}
\end{array}\right]\equiv \left[\begin{array}{cc}
	-\delta^2k^2+1-3U_h^2 & -1 \\ \varepsilon & -(k^2+\varepsilon\alpha)
\end{array}\right].
$$
The solvability condition at this order in $\epsilon$ 
$${\rm det}\left(L_1^{(1)}-\sigma\mathbf{I}_{2\times2}\right)=0,$$
then leads to a quadratic equation in $\sigma$
\begin{equation}\label{disp_re}
	\sigma^2-T_1(k)\sigma+\Delta_1(k)=0,
\end{equation}
with 
\begin{equation}
	T_1(k)\equiv {\rm tr}L^{(1)}=L_{11}^{(1)}+L_{22}^{(1)},
\end{equation}
and 
\begin{equation}
	\Delta_1(k)\equiv {\rm det}L^{(1)}=L_{11}^{(1)}L_{22}^{(1)}+\varepsilon.
\end{equation}
The solution of Eq.~(\ref{disp_re}) corresponds to the dispersion relation between the growth rate $\sigma$ and the wavenumber $k$ shown in Eq.~(\ref{disperelation}).
\section{Weakly nonlinear analysis around the Turing bifurcation}\label{append2}
	Here $U_h=U_T=$ (i.e., $b=b_T$) and the appropriate asymptotic expansion for the variables previously defined is
\begin{equation}\label{eq.HSS}
	\left[\begin{array}{c}
		U_h\\V_h
	\end{array}\right]=\left[\begin{array}{c}
		U_T\\V_T
	\end{array}\right]+ \epsilon^2\left[\begin{array}{c}
		U_2\\V_2
	\end{array}\right]+...
\end{equation}
\begin{equation}
	\left[\begin{array}{c}
		\phi\\\psi
	\end{array}\right]= \epsilon\left[\begin{array}{c}
		\phi_1\\\psi_1
	\end{array}\right]+ \epsilon^2\left[\begin{array}{c}
		\phi_2\\\psi_2
	\end{array}\right]+ \epsilon^3\left[\begin{array}{c}
		\phi_3\\\psi_3
	\end{array}\right]+...,
\end{equation}
where we allow all the variables $\phi_1,\psi_1,\phi_2,\psi_2,...$ to be functions of $x$ and the long spatial scale $X\equiv\epsilon x$ [e.g., $=\phi_1(x,X=\epsilon x)$]
	\begin{equation}
	\mathcal{N}\equiv -u^2\mathcal{M}\equiv- u^2 \left[\begin{array}{cc}
		1 & 0  \\ 0 & 0
	\end{array}\right]
\end{equation}

{\bf B.1. The homogeneous problem}\\\\
Before proceeding further, we consider the uniform problem, which satisfies: 
\begin{equation}
	\left(\mathcal{L}^h+\mathcal{N}^h\right) \left[\begin{array}{c}
		U_h \\ V_h
	\end{array}
	\right]+\mathcal{Y}=\left[\begin{array}{c}
		0 \\ 0
	\end{array}
	\right],
\end{equation}
with 
\begin{equation}
	\mathcal{L}^h=\mathcal{B}, \qquad \mathcal{N}^h\equiv- U_h^2 \mathcal{M},
\end{equation}
Considering the expansion 
\begin{equation}\label{eq.HSS}	\left[\begin{array}{c}
		U_h\\V_h
	\end{array}\right]= \left[\begin{array}{c}
		U_T\\V_T
	\end{array}\right]+ \epsilon^2\left[\begin{array}{c}
		U_2\\V_2
	\end{array}\right]+...
\end{equation}
nonlinear operators expand as  $\mathcal{N}^h=\mathcal{N}_0^h+\epsilon^2\mathcal{N}_2^h+\mathcal{O}(\epsilon^3)$, with 
\begin{equation}
	\mathcal{N}_0^h\equiv -U_T^2\mathcal{M},\qquad
	\mathcal{N}_2^h\equiv -2U_TU_2\mathcal{M},  
\end{equation}
the term $\mathcal{Y}$ becomes
\begin{equation}
	\mathcal{Y}=\mathcal{Y}_2+\epsilon^2\mathcal{Y}_2  \equiv \left[\begin{array}{c}
		-\beta_T \\ 0
	\end{array}
	\right]+\epsilon^2\left[\begin{array}{c}
		-\mu \\ 0
	\end{array}
	\right],
\end{equation}
and the linear operator is $\mathcal{O}(\epsilon^0)$, and thus $\mathcal{L}^h=\mathcal{L}_0^h$. 

The uniform equation splits order by order as follows:

At order $\epsilon^0$, 
we have the equation
$$
\mathcal{O}(\epsilon^0):\quad	\left(\mathcal{L}_0^h+\mathcal{N}_0^h\right) \vec{Q}_T+\mathcal{Y}_T=\vec{0},
$$
whose solution is the one obtained in Section~\ref{sec:2}.

At order $\epsilon^2$, 
we get the equation
$$
\mathcal{O}(\epsilon^2):\quad	\left(\mathcal{L}_0^h+\mathcal{N}_0^h\right) \vec{Q}_2+\mathcal{N}_2^h\vec{Q}_0+\mathcal{Y}_2=\vec{0},
$$
and if we use the equality 
$U_2\mathcal{M}\vec{Q}_0=u_T\mathcal{M}\vec{Q}_2,$
we finally get 
\begin{equation}\label{hom_2nd_order}
L_0\vec{Q}_2+\mathcal{Y}_2=0,
\end{equation}
with 
$$L_0\equiv\left(\mathcal{L}_0^h+3\mathcal{N}_0^h\right)=\left[\begin{array}{cc}
	L_{11}^{(0)} & -1 \\ \varepsilon & L_{22}^{(0)}
\end{array}\right]=\left[\begin{array}{cc}
	1-3u_T^2 & -1 \\ \varepsilon & -\varepsilon\alpha
\end{array}\right]$$
The solution of Eq.~(\ref{hom_2nd_order}) is given by 
\begin{equation}
	\vec{Q}_2=-L_0^{-1}\mathcal{Y}_2,
\end{equation}
with 
$$L_0^{-1}=\frac{1}{\Delta_0}\left[\begin{array}{cc}
	L_{11}^{(0)} & -\varepsilon \\ 1 & 	L_{22}^{(0)}
\end{array}\right],\quad\Delta_0\equiv L_{11}^{(0)}L_{22}^{(0)}+\varepsilon.$$
After some algebra, we obtain
\begin{equation}
	\vec{Q}_2=	\left[\begin{array}{c}
		U_2 \\ V_2
	\end{array}
	\right]=\mu\vec{G}_2, \quad \vec{G}_2\equiv	\left[\begin{array}{c}
		G_2^{(1)} \\ G_2^{(2)}
	\end{array}
	\right]\equiv\frac{1}{\Delta_0}	\left[\begin{array}{c}
		L_{11}^{(0)} \\ 1
	\end{array}
	\right],
\end{equation}
which is the solution we were looking for.
\\\\
{\bf B.2. The spatially-dependent problem}\\\\
We now consider the full problem, including the coupling between uniform and space-dependent components. The linear and nonlinear operators are expanded as:

$$\mathcal{L}=\mathcal{L}_0+ \epsilon\mathcal{L}_1+\epsilon^2\mathcal{L}_2+\mathcal{O}(\epsilon^3),$$
with 
\begin{equation}
\mathcal{L}_0\equiv\mathcal{L}_0^h+\mathcal{A}\partial_x^2,\qquad\mathcal{A}\equiv\left[\begin{array}{cc}
	\delta^2 & 0  \\ 0 & 1
\end{array}\right],
\end{equation}
\begin{equation}
	\mathcal{L}_1\equiv2\mathcal{A}\partial_x\partial_X,\qquad\mathcal{L}_2\equiv\mathcal{A}\partial_X^2,
\end{equation}
and
$$\mathcal{N}=\mathcal{N}_0+\mathcal{N}_1\epsilon+\epsilon^2\mathcal{N}_2+\epsilon^3\mathcal{N}_3+\mathcal{O}(\epsilon^4),$$
with the terms
$$\mathcal{N}_0\equiv\mathcal{N}_0^h\qquad\mathcal{N}_1\equiv -2 U_T \phi_1\mathcal{M}, $$
$$\mathcal{N}_2\equiv\mathcal{N}_2^h+\mathcal{N}_2^s,\quad\mathcal{N}_2^s\equiv-(\phi_1^2+2U_T\phi_2)\mathcal{M},$$
$$\mathcal{N}_3\equiv-2(\phi_1\phi_2+U_T\phi_3+U_2\phi_1)\mathcal{M}.$$
Thus, order by order we obtain:
$$
\begin{array}{ll}
	\mathcal{O}(\epsilon^0):& \left(\mathcal{L}_0+\mathcal{N}_0\right)Q_0+\mathcal{Y}_0=0\\
	\mathcal{O}(\epsilon^1):& \left(\mathcal{L}_0+\mathcal{N}_0\right)q_1+\mathcal{N}_1Q_0=0,\\
	\mathcal{O}(\epsilon^2):& \left(\mathcal{L}_0+\mathcal{N}_0\right)q_2+\left(\mathcal{L}_1+\mathcal{N}_1\right)q_1+\mathcal{N}_2Q_0+\mathcal{Y}_2=0\\
	\mathcal{O}(\epsilon^3):& \left(\mathcal{L}_0+\mathcal{N}_0\right)q_3+\left(\mathcal{L}_1+\mathcal{N}_1\right)q_2+\left(\mathcal{L}_2+\mathcal{N}_2\right)q_1+\mathcal{N}_3Q_0=0.
\end{array}
$$
\\\\
{\bf The spatial-dependent solution at $\mathcal{O}(\epsilon)$}\\\\
At first order in $\epsilon$, the equation 
$$\left(\mathcal{L}_0+\mathcal{N}_0\right)q_1+\mathcal{N}_1Q_0=\vec{0}$$
becomes 
\begin{equation}\label{WNL_1}
	Lq_1\equiv\left(\mathcal{L}_0+3\mathcal{N}_0^h\right)q_1=\vec{0},
\end{equation}
where we have used the equality 
\begin{equation}\label{relation}
	q^{(m)}\mathcal{M}Q=Q^{(m)}\mathcal{M}q,
\end{equation}
being $q$ and $Q$ any bidimensional vectors. In full matrix form, this linear operator reads 
$$
L\equiv\left[\begin{array}{cc}
	L_{11} & L_{12} \\ L_{21} & L_{22}
\end{array}\right]\equiv \left[\begin{array}{cc}
	\delta^2\partial_x^2+1-3U_T^2 & -1 \\ \varepsilon & \partial_x^2-\varepsilon\alpha
\end{array}\right].	
$$
As solution for Eq.~(\ref{WNL_1}) we propose the ansatz
\begin{equation}\label{ansatz}
	\vec{q}_1=\vec{\xi}\left(A(X)e^{i k_c x}+c.c.\right),\quad \vec{\xi}=\left[\begin{array}{c}
		\xi_1 \\ \xi_2
	\end{array}
	\right]
\end{equation}
Applying $L$ to the previous ansatz we obtain
$L_1\vec{\xi}=0$, with 
$$
L_1\equiv\left[\begin{array}{cc}
	L_{11}^{(1)} & -1 \\ \varepsilon & L_{22}^{(1)}
\end{array}\right]\equiv \left[\begin{array}{cc}
	-\delta^2k_T^2+1-3U_T^2 & -1 \\ \varepsilon & -k_T^2-\varepsilon\alpha
\end{array}\right].	
$$
With ansatz, $L_1\vec{\xi}=0$ has non-trivial solutions if $\Delta_1\equiv{\rm det}[L_1]=L_{11}^{(1)}L_{22}^{(1)}+\varepsilon=0$, which leads to the condition
\begin{equation}\label{solv_1}
(\delta^2k^2+3U_T^2-1)(k^2+\varepsilon\alpha)+\varepsilon=0,
\end{equation}
is satisfied. This leads to components $\xi=1$ and $\xi_2=L_{11}^{(1)}$ in the ansatz (\ref{ansatz}).
\\\\
{\bf The spatially-dependent solution at $\mathcal{O}(\epsilon^2)$}\\\\
Reorganizing the equation we obtained at this order we have 
$$\underbrace{\left(\mathcal{L}_0+\mathcal{N}_0^h\right)Q_2+\mathcal{N}_2^hQ_0+\mathcal{Y}_2}_{\displaystyle\vec{0}}+$$$$\left(\mathcal{L}_0+\mathcal{N}_0\right)q_2+\left(\mathcal{L}_1+\mathcal{N}_1\right)q_1+\mathcal{N}_2^sQ_0=0,$$
and therefore we have 
\begin{equation}\label{eq_orden2}
\left(\mathcal{L}_0+\mathcal{N}_0\right)q_2+\left(\mathcal{L}_1+\mathcal{N}_1\right)q_1+\mathcal{N}_2^sQ_0=0.
\end{equation}
If we use Eq.~(\ref{relation}), we can write 
$$\mathcal{N}_2^sQ_0=-2U_T^2\mathcal{M}q_2-\phi_1U_T\mathcal{M}q_1,$$
and thus Eq.~(\ref{eq_orden2}) eventually becomes 
\begin{equation}\label{eq_order2}
	Lq_2=
	-\left(\mathcal{L}_1+\frac{3}{2}\mathcal{N}_1\right)q_1.
\end{equation}
To continue from here, we must compute the RHS of Eq.~(\ref{eq_order2}). First, 
$$\left(\mathcal{L}_1+\frac{3}{2}\mathcal{N}_1\right)q_1=\left(2\mathcal{A}\partial_x\partial_X-3U_T\phi_1\mathcal{M}\right)\left[\begin{array}{c}
	\phi_1\\\psi_1
\end{array}\right]=$$
$$2\left[\begin{array}{c}
	\delta^2\\L_{11}^{(1)}
\end{array}\right](ik_T\partial_XA e^{ik_Tx}+c.c.)-3U_T\left[\begin{array}{c}
1\\0
\end{array}\right]\left(2|A|^2+A^2e^{2ik_Tx}+c.c.\right)=$$
$$\vec{f}_0|A|^2+i\vec{f}_1A_Xe^{ik_Tx}+\vec{f}_2A^2e^{2ik_Tx}+c.c.,$$
with 
$$\vec{f}_0\equiv-6U_T\left[\begin{array}{c}
	1\\0
\end{array}\right],\quad \vec{f}_1\equiv2k_T\left[\begin{array}{c}
\delta^2\\L_{11}^{(1)}
\end{array}\right],\quad \vec{f}_2\equiv f_0/2.$$
At this stage, we need to apply the F alternative and derive a solvability condition at this order in $\epsilon$.
To do so we first define the scalar product
$$\langle \vec{f}|\vec{g}\rangle=\frac{1}{l}\int_{-l/2}^{l/2} \vec{f}^T(x)\cdot \vec{g}(x)dx.$$
where $l=2\pi/k_T$. With this definition and the adjoint operator associated with $L$, namely 
$$L_1^\dagger=\left[\begin{array}{cc}
	L_{11}^{(1)} & \varepsilon \\ -1 & L_{22}^{(1)}
\end{array}\right],$$
we can write 
\begin{equation}\label{solv_cond2}
	\langle\vec{w}|L\vec{f}\rangle=\langle L^\dagger\vec{w}|\vec{f}\rangle=0,
\end{equation}
where
$$\vec{w}=\vec{W}e^{ik_Tx}+c.c.,\quad \vec{W}\equiv\left[\begin{array}{c}
	w_1\\w_2
\end{array}\right]=\left[\begin{array}{c}
L_{22}^{(1)}\\1
\end{array}\right]$$
is the null-vector associated with $L^\dagger$,i.e., $L^\dagger\vec{w}=\vec{0}$.

The solvability condition associated with Eq.~(\ref{eq_order2}) then reads 
\begin{equation}\label{solv_cond2}
\langle\vec{w}|\left(\mathcal{L}_1+3\mathcal{N}_1/2\right)q_1\rangle=0,
\end{equation}
After some algebra, Eq.~(\ref{solv_cond2}) yields:
\begin{equation}
	k_T(\delta^2L_{22}^{(1)}+L_{11}^{(1)})=0.
\end{equation}
For $K_T\neq0$, this condition leads to 
\begin{equation}\label{solv_2}
	\delta^2k^2+3U_T^2-1=-\delta^2(k^2+\varepsilon\alpha).
\end{equation}
At this point we must make a brake in our derivation to point out that combining Eq.~(\ref{solv_1}) and Eq.~(\ref{solv_2}) we obtain the critical wavenumber associated to the Turing bifurcation 
	\begin{equation}\label{kc}
	k_T=\sqrt{-\varepsilon \alpha+\sqrt{\varepsilon}/\delta},
\end{equation}
provided that $\delta<1/(\alpha\sqrt{\epsilon}),$
and its position
\begin{equation}
	U^{\pm}_T=\pm\sqrt{\frac{1+\delta^2\varepsilon\alpha-2\delta\sqrt{\varepsilon}}{3}}, \end{equation}
as already derived in Section~\ref{sec:3}.

When the solvability condition (\ref{solv_2}) is satisfied, a proper ansatz to solve Eq.~(\ref{eq_order2}) reads
$$
\left[\begin{array}{c}
 		\phi_2\\\psi_2
 	\end{array}\right]=\underbrace{\left[\begin{array}{c}
 	a_1\\a_2
 \end{array}\right]}_{\displaystyle\vec{a}}|A|^2+\underbrace{\left[\begin{array}{c}
 b_1\\b_2
\end{array}\right]}_{\displaystyle\vec{b}}iA_Xe^{ik_Tx}+\underbrace{\left[\begin{array}{c}
c_1\\c_2
\end{array}\right]}_{\displaystyle\vec{c}}A^2e^{2ik_Tx}+c.c.
$$
If we collect all the terms multiplying the same exponential we obtain three equations:
\begin{equation}
L_0\left[\begin{array}{c}
	a_1\\a_2
\end{array}\right]=f_0,\quad L_1\left[\begin{array}{c}
b_1\\b_2
\end{array}\right]=f_1, \quad L_2\left[\begin{array}{c}
c_1\\c_2
\end{array}\right]=f_2,
\end{equation}
where 
$$L_0\equiv\left[\begin{array}{cc}
	L_{11}^{(0)} & -1 \\ \varepsilon & L_{22}^{(0)}
\end{array}\right]\equiv \left[\begin{array}{cc}
	1-3U_T^2 & -1 \\ \varepsilon & -\varepsilon\alpha
\end{array}\right],$$

$$L_2\equiv\left[\begin{array}{cc}
	L_{11}^{(2)} & -1 \\ \varepsilon & L_{22}^{(2)}
\end{array}\right]\equiv \left[\begin{array}{cc}
	-4k_T^2\delta^2+1-3U_T^2 & -1 \\ \varepsilon & -4k_T^2-\varepsilon\alpha
\end{array}\right].$$
The solutions of the previous first and third equations read
$$
\left[\begin{array}{c}
	a_1\\a_2
\end{array}\right]=L_0^{-1}f_0=\frac{-6U_T}{\Delta_0}\left[\begin{array}{cc}
L_{11}^{(0)} & -\varepsilon \\ 1 & L_{22}^{(0)}
\end{array}\right]\left[\begin{array}{c}
1\\0
\end{array}\right],
$$
$$
\left[\begin{array}{c}
	c_1\\c_2
\end{array}\right]=L_2^{-1}f_2=\frac{-3U_T}{\Delta_2}\left[\begin{array}{cc}
	L_{11}^{(2)} & -\varepsilon \\ 1 & L_{22}^{(2)}
\end{array}\right]\left[\begin{array}{c}
	1\\0
\end{array}\right],
$$
with the determinant associated with the linear operator $L_m$ given by $\Delta_m=L_{11}^{(m)}L_{22}^{(m)}+\varepsilon$, for $m=0,1,2$.

For the second equation, $\Delta_1=0$, and therefore we cannot solve it in the same way, because $L_1$ is not invertible. To solve this equation we proceed in a different way. Combining both equation we obtain the following expressions
\begin{equation}
b_1\Delta_1=2k_T(L_{11}^{(1)}+\delta^2L_{22}^{(1)}),\qquad b_2=L_{11}^{(1)}b_1-2k_T\delta^2.
\end{equation} 
The right-hand side of the first equation is zero as the solvability condition is $\Delta_1$. This implies that $b_1$ can take any value, and without loss of generality we can choose $b_1=0$. With this we obtain
\begin{equation}
\left[\begin{array}{c}
	b_1\\b_2
\end{array}\right]=\left[\begin{array}{c}
0\\-2k_T\delta^2
\end{array}\right].
\end{equation}


{\bf The spatially-dependent solution at $\mathcal{O}(\epsilon^3)$}\\\\
Following similar steps, we need to simplify equation at $\mathcal{O}(\epsilon^3)$. The first thing to do is to rewrite the last term $\mathcal{N}_3Q_0$ using the relation (\ref{relation}), which yields 
$$\mathcal{N}_3Q_0=-2U_T^2\mathcal{M}q_3-2U_2U_T\mathcal{M}q_1-2U_T\phi_1\mathcal{M}q_2.$$
With this expression, our equation at this order becomes 
$$	Lq_3=-\left(\mathcal{L}_1+\mathcal{N}_1-2U_T\phi_1\mathcal{M}\right)q_2-\left(\mathcal{L}_2+\mathcal{N}_2-2U_TU_2\mathcal{M}\right)q_1=$$
\begin{equation}\label{3rd_order}
L\vec{q}_3=-\left(\mathcal{L}_1+2\mathcal{N}_1\right)\vec{q}_2-(\mathcal{L}_2+2\mathcal{N}_2^h+\mathcal{N}_2^s)\vec{q}_1
\end{equation}
 
First, the second term in Eq.~(\ref{3rd_order}) becomes
\begin{equation}
	(\mathcal{L}_1+2\mathcal{N}_1)\vec{q}_2=g_0+g_1e^{ik_cx}+g_2e^{2ik_cx}+g_3e^{3ik_cx},
\end{equation}
with 
$$	g_1=g_1^bA_{XX}+g_1^c|A|^2A,$$
and 
\begin{equation}
g_1^b\equiv-2k_T\mathcal{A}\vec{b}, \quad g_1^c\equiv-4U_T\mathcal{M}(\vec{a}+\vec{c}).
\end{equation}

The third term becomes
\begin{equation}
	(\mathcal{L}_2+2\mathcal{N}^h_2+\mathcal{N}_2^s)\vec{q}_1=h_0+h_1e^{ik_cx}+h_2e^{2ik_cx}+h_3e^{3ik_cx},
\end{equation}
with 
\begin{equation}
	h_1=h_1^aA+h_1^bA_{XX}+h_1^c|A|^2A,
\end{equation}
and
	\begin{equation}
		h_1^a=-4U_TU_2\mathcal{M}\vec{\xi}=-4\mu U_TG_2^{(1)}\mathcal{M}\vec{\xi}=\mu H_1^a,
	\end{equation}
	\begin{equation}
		h_1^b=\mathcal{A}\vec{\xi},
	\end{equation}
	\begin{equation}
		h_1^c=-\left(2U_T(a_1+c_1)+3\right)\mathcal{M}\vec{\xi}.
	\end{equation}
The solvability condition 
$$
	\langle \vec{w}|(\mathcal{L}_1+2\mathcal{N}_1)\vec{q}_2\rangle+\langle \vec{w}|(\mathcal{L}_2+2\mathcal{N}^h_2+\mathcal{N}_2^s)\vec{q}_1\rangle=0,
$$
leads to 
$$\mu\vec{W}^T \vec{H}_1^aA+\vec{W}^T(\vec{g}_1^b+\vec{h}_1^b)A_{XX}+\vec{W}^T(\vec{g}_1^c+\vec{h}_1^c)|A|^2A=0$$
with 
$$\vec{W}^T \vec{H}_1^a=-4 U_TG_2^{(1)}\vec{W}^T\mathcal{M}\vec{\xi},$$
$$\vec{W}^T (\vec{g}_1^a+\vec{h}_1^b)=-2k_T\vec{W}^T\mathcal{A}\vec{b}+\vec{W}^T\mathcal{A}\vec{\xi},$$
$$\vec{W}^T (\vec{g}_1^c+\vec{h}_1^c)=-\left(6U_T(a_1+c_1)+3\right)\vec{W}^T\mathcal{M}\vec{\xi}.$$
At this point we only have to perform the vector/matrix multiplication, which leads to 
$$\vec{W}^T\mathcal{M}\vec{\xi}=\left[\begin{array}{cc}
L_{22}^{(1)}& 1
\end{array}\right]\left[\begin{array}{cc}
1& 0\\0&0
\end{array}\right]\left[\begin{array}{c}
1\\ L_{11}^{(1)}
\end{array}\right]=L_{22}^{(1)},$$

$$\vec{W}^T\mathcal{A}\vec{\xi}=\left[\begin{array}{cc}
	L_{22}^{(1)}& 1
\end{array}\right]\left[\begin{array}{cc}
	\delta^2& 0\\0&1
\end{array}\right]\left[\begin{array}{c}
	1\\ L_{11}^{(1)}
\end{array}\right]=L_{22}^{(1)}\delta^2+L_{11}^{(1)}=0,$$

$$\vec{W}^T\mathcal{A}\vec{b}=\left[\begin{array}{cc}
	L_{22}^{(1)}& 1
\end{array}\right]\left[\begin{array}{cc}
	\delta^2& 0\\0&1
\end{array}\right]\left[\begin{array}{c}
	b_1 \\ b_2
\end{array}\right]=L_{22}^{(1)}\delta^2b_1+b_2.$$

\begin{figure}[!t]
\centering
\includegraphics[scale=1.0]{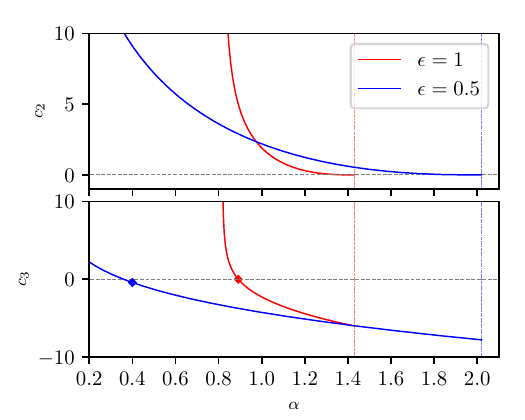}  
		\caption{Modification of the normal form coefficients as a function of $\alpha$ for $\delta=0.7$ and different values of $\epsilon$.}
		\label{wnl_coeff}
	\end{figure}
After all this we finally get the amplitude equation 
\begin{equation}\label{amplitude_eq0}
	\mu A+C_2A_{XX}+C_3A|A|^2=0,
\end{equation}
with the coefficients 
\begin{equation}
	C_2\equiv\frac{\vec{W}^T(\vec{g}_1^b+\vec{h}_1^b)}{\vec{W}^T \vec{H}_1^a}=\frac{k_T}{2U_T}\frac{L_{22}^{(1)}\delta^2b_1+b_2}{G_2^{(1)}L_{22}^{(1)}},
\end{equation}
\begin{equation}
	C_3\equiv\frac{\vec{W}^T(\vec{g}_1^c+\vec{h}_1^c)}{\vec{W}^T \vec{H}_1^a}=\frac{6U_T(a_1+c_1)+3}{4U_TG_{2}^{(1)}}.
\end{equation}
Our next step is to solve the amplitude equation.
\\\\
{\bf Solution of the amplitude equation}\\\\
The amplitude Eq.~(\ref{amplitude_eq0}) has two type of solutions: extended ones and localized ones. Let us first take $A(X)=Z(X)e^{i\varphi}$, where we have consider that phase does not depend on $X$. This yields:
 \begin{equation}\label{amplitude_eq00}
 	\mu Z+C_2Z_{XX}+C_3Z^3=0.
 \end{equation}
If $Z\neq Z(X)$, i.e., the amplitude of the solution is constant in $X$, the previous equation becomes 
 \begin{equation}\label{amplitude_eq000}
	\mu Z+C_3Z^3=0,
\end{equation}
which is the normal form of a Pitchfork bifurcation \cite{wiggins_introduction_2003}. This equation supports the solutions $Z=0$, which exist always, and 
\begin{equation}
	Z=\sqrt{-\mu/C_3}.
\end{equation}
Depending on the value of the sign of $C_3$, this solution will exist for $\mu<0$ if $C_3>0$, or for $\mu>0$, if $C_3<0$. In the first case we are in a sub-critical pitchfork regime, while in the second the pitchfork is supercritical. The transition between these two situations occurs for $C_3=0$. A general analytical solution of this equation is not tractable. However, we can solve this equation numerically. Figure~\ref{wnl_coeff}(a) shows the dependence of $C_3$ on $\alpha$ for $\delta=0.7$ and two different values of $\epsilon$, which corresponds to the phase diagrams shown in Figs.~\ref{fig_phase_dia_1} and \ref{fig_phase_dia_2}(a). The intersection with the horizontal line at zero provides the $\alpha$-value of the codimension-two point where the pattern changes its criticality. Besides, these curves extend until the position marker with a point-dashed vertical line, which signals the BD point for each configuration. 

If we allow $Z$ to depend on $X$, i.e., $Z=Z(X)$, Eq.~(\ref{amplitude_eq00}) supports localized solutions 
\begin{equation}
	Z(X)=\sqrt{\frac{-2\mu}{C_3}}{\rm sech}\left(\sqrt{\frac{-\mu}{C_1}}X\right),
\end{equation}
provided that $C_3>0$.

\bibliographystyle{ieeetr}
\bibliography{Refs}
\end{document}